\documentclass[final,3p,times]{elsarticle}
\pdfoutput=1
\usepackage{bbding}
\usepackage{pifont}
\usepackage{amsfonts}
\usepackage{mathrsfs}
\usepackage{amsthm}
\usepackage{graphicx}
\usepackage{subfig}
\usepackage{epstopdf}
\usepackage[tbtags]{amsmath}
\usepackage{mathtools}
\usepackage{color}
\usepackage{arydshln}
\usepackage{amssymb}
\usepackage{booktabs}
\usepackage{bookmark}
\usepackage{tikz}
\usepackage{multirow}

\newcommand*{\circled}[1]{\lower.7ex\hbox{\tikz\draw (0pt, 0pt)%
		circle (.5em) node {\makebox[1em][c]{\small #1}};}} %圆圈数字
\newcommand{\Hinf}{${H}_{\infty}\ $}

\newcommand{\Ltwo}{${\mathcal{L}}_{2}\ $}

\newtheorem{theorem}{Theorem}
\newtheorem{lemma}{Lemma}

\newdefinition{remark}{Remark}
\newdefinition{definition}{Definition}
\newproof{Proof}{\bf{Proof}}

\biboptions{sort&compress}

\journal{ }

\begin{document}
	
	\begin{frontmatter}
		
		%% Title, authors and addresses
		
		%% use the tnoteref command within \title for footnotes;
		%% use the tnotetext command for theassociated footnote;
		%% use the fnref command within \author or \address for footnotes;
		%% use the fntext command for theassociated footnote;
		%% use the corref command within \author for corresponding author footnotes;
		%% use the cortext command for theassociated footnote;
		%% use the ead command for the email address,
		%% and the form \ead[url] for the home page:
		%% \title{Title\tnoteref{label1}}
		%% \tnotetext[label1]{}
		\author[1]{Jiamin Wang}
		\ead{wangjiamin21@outlook.com}
		\author[1]{Liqi Zhou}
		\ead{lqzhou\_96@foxmail.com}
		\author[2]{Dong Zhang}
		\ead{zhangdong@nwpu.edu.cn}
		\author[1]{Jian Liu\corref{cor1}}
		\ead{liujianzym@outlook.com}
		\author[1]{Yuanshi Zheng\corref{cor1}}
		\ead{zhengyuanshi2005@163.com}
		\affiliation[1]{
			organization={Shaanxi Key Laboratory of Space Solar Power Station System, School of Mechano-electronic Engineering, Xidian University},
			%	addressline={},
			city={Xi'an},
			postcode={710071},
			%	state={shannxi},
			country={China}}
		\affiliation[2]{
			organization={School of Astronautics, Northwestern Polytechnical University},
			%	addressline={},
			city={Xi'an},
			postcode={710072},
			%	state={shannxi},
			country={China}}
		% \affiliation[2]{
			% 	organization={School of Astronautics, Northwestern Polytechnical University},
			%	%	addressline={},
			%	city={Xi'an},
			%	postcode={710072},
			%	%	state={shannxi},
			%	country={China}}
		% \ead[url]{home page}
		% \fntext[label2]{}
		\cortext[cor1]{Corresponding authors}

		\title{Protocol selection for second-order consensus against disturbance \tnoteref{t1}}
		\tnotetext[t1]{This work was supported in part by the National Natural Science Foundation of China under Grant 62273267; in part by the Natural Science Basic Research Program of Shaanxi under Grant 2022JC-46.}
		%% use optional labels to link authors explicitly to addresses:
		%% \author[label1,label2]{}
		%% \affiliation[label1]{organization={},
			%%             addressline={},
			%%             city={},
			%%             postcode={},
			%%             state={},
			%%             country={}}
		%%
		%% \affiliation[label2]{organization={},
			%%             addressline={},
			%%             city={},
			%%             postcode={},
			%%             state={},
			%%             country={}}
		
		%\author{}
		%
		%\affiliation{organization={},%Department and Organization
			%            addressline={}, 
			%            city={},
			%            postcode={}, 
			%            state={},
			%            country={}}
		
		\begin{abstract}
			%% Text of abstract
			Noticing that both the absolute and relative velocity protocols can solve the second-order consensus of multi-agent systems, this paper aims to investigate which of the above two protocols has better anti-disturbance capability, in which the anti-disturbance capability is measured by the \Ltwo gain from the disturbance to the consensus error. More specifically, by the orthogonal transformation technique, the analytic expression of the \Ltwo gain of the second-order multi-agent system with absolute velocity protocol is firstly derived, followed by the counterpart with relative velocity protocol. It is shown that both the \Ltwo gains for absolute and relative velocity protocols are determined only by the minimum non-zero eigenvalue of Laplacian matrix and the tunable gains of the state and velocity. 
			Then, we establish the graph conditions to tell which protocol has better anti-disturbance capability. Moreover, we propose a two-step scheme to improve the anti-disturbance capability of second-order multi-agent systems. Finally, simulations are given to illustrate the effectiveness of our findings.
		\end{abstract}
		
		%%%Graphical abstract
		%\begin{graphicalabstract}
		%%\includegraphics{grabs}
		%\end{graphicalabstract}
		%
		%%%Research highlights
		%\begin{highlights}
		%\item Research highlight 1
		%\item Research highlight 2
		%\end{highlights}
		
		\begin{keyword}
			%% keywords here, in the form: keyword \sep keyword
			%% PACS codes here, in the form: \PACS code \sep code
			%% MSC codes here, in the form: \MSC code \sep code
			%% or \MSC[2008] code \sep code (2000 is the default)
			Multi-agent systems \sep Second-order consensus \sep Anti-disturbance capability \sep Graph condition \sep Protocol selection
		\end{keyword}
		
	\end{frontmatter}
	
	%% \linenumbers
	
	%% main text
	\section{Introduction}\label{Sec:1}
	Over the past decades, the distributed coordination of multi-agent systems (MASs) has been extensively investigated in control community.
	As a fundamental issue in multi-agent coordination, consensus problem has attracted tremendous attention. 
	Roughly speaking, consensus means that a group of agents reaches an agreement regarding a common quantity of interest by designing appropriate communication protocol \cite{Olfati2007}.
	
	In retrospect, for the consensus problem, agents are assumed to took first-order dynamics in early seminal works \cite{Jadbabaie2003,Olfati2004,Ren2005}.
	However, the first-order dynamics model is hardly capable of describing many mechanical systems. Therefore, numerous researchers started to investigate the consensus protocols for second-order MASs to overcome this shortcoming.
	Subsequently, two classic second-order consensus protocols emerged which employed different treatments to velocity information.
	One protocol proposed in \cite{Xie2007} introduced the absolute velocity information of the agent itself as the local feedback. Another protocol devised in \cite{Ren2007} required each agent to use the relative velocity measurements with respect to its neighbors.
	Then a series of researches on second-order consensus sprang up.
	More general forms of second-order consensus protocols were studied in \cite{Zhu2009,Yu2010,Mei2015}.
	Some results took the communication delay into consideration \cite{Lin2009,Qin2011,Hou2017}.
	%It is worth mentioning that the inertias and control gains of agents were considered to be heterogeneous in \cite{Mei2015}.
	The authors in \cite{Ai2016} addressed the consensus of second-order MASs with limited agent interaction ranges.
	In order to reduce the resource consumption, the event-based consensus protocol was developed for second-order MASs in \cite{Zhu2017}.
	The resilient consensus was studied in \cite{Dibaji2017} for second-order MASs with faulty or malicious agents.
	The authors in \cite{Zheng2011,Zheng2019,Zhao2020-1,ZhuYR2020} investigated the consensus of heterogeneous and hybrid MASs, in which agents have different dynamics behaviours.
	
	Since the disturbance is rife in reality, it is of great significance to investigate the anti-disturbance capability for second-order MASs with absolute or relative velocity protocol. 
	In literature, the \Ltwo gain is popular in MAS community to characterize the influence of disturbance on the consensus. 
	For example, by LMIs technique, the \Ltwo gain optimal consensus is considered in \cite{Lin2010,Li2015,Han2016,Huang2018,Lin2008,Li2011,Liu2012,Wang2014}.
	In a separate direction, some studies tapped into the role of networks on the \Ltwo gain of MASs \cite{Siami2014,Pirani2018,Pirani2019,Yang2012}.
	In \cite{Siami2014}, the authors built the relation between the \Ltwo gain of first-order MASs and the minimum non-zero eigenvalue of the Laplacian matrix associated with the undirected graph.
	As pointed out in \cite{Pirani2018}, the \Ltwo gain of first-order leader-follower MASs on undirected graphs relies on the minimum eigenvalue of the grounded Laplacian matrix. For first-order leader-follower MASs on directed graphs, it was shown in \cite{Pirani2019} that the \Ltwo gain of first-order leader-follower MASs on directed graphs depends on the minimum singular value of the grounded Laplacian matrix.
	Besides, the authors in \cite{Yang2012} analyzed the \Ltwo gain of second-order leader-follower MASs with the relative velocity protocol in the presence of communication errors and measurement errors, and derived the analytic expressions of \Ltwo gain for a class of special directed networks in which the in-degree of each node was assumed to be the same.
	Note that the second-order consensus protocols in all aforementioned literature were founded on the basic structures of two classic absolute velocity protocol \cite{Xie2007} and relative velocity protocol \cite{Ren2007}.
	Therefore, it is of great significance to study the analytic expressions of \Ltwo gains for the second-order MASs with general absolute and relative velocity protocols and develop the graph conditions to tell which one of the above two protocols has better anti-disturbance capability for the second-order consensus.
	As far as we known, no previous study has investigated these issues.
	
	Motivated by the above observations, this paper aims to investigate which one of the absolute and relative velocity protocols has better anti-disturbance capability for consensus of second-order MASs.
	The considered problem is challenging as the general communication topology results in difficulty to establish the quantitative relations between weighted adjacency matrix, tunable gains and the anti-disturbance capability.
	In this paper, the anti-disturbance capability is measured by the \Ltwo gain from the disturbance to the consensus error, and we intend to establish the quantitative relations between weighted adjacency matrix, tunable gains and anti-disturbance capability for the second-order MASs with absolute and relative velocity protocols, respectively.
	Furthermore, on the basis of the established quantitative relations, we give the graph conditions of protocol selection for better anti-disturbance capability.
	Our contributions are summarized as follows:
	\begin{itemize}
		\item[1)] By the orthogonal transformation technique, we establish the quantitative relations between weighted adjacency matrix, tunable gains and the \Ltwo gains for the second-order MASs with absolute and relative velocity protocols, respectively. It is shown that the \Ltwo gains are inversely proportional to both the minimum non-zero eigenvalue of the Laplacian matrix and the tunable gains.
		\item[2)] The protocol selection criteria are developed for the second-order MASs. We give the graph conditions to tell which one of the absolute and relative velocity protocols has better anti-disturbance capability.
		%	We give the graph conditions to tell which one of the absolute and relative velocity protocols has better anti-disturbance capability. It is proved that the protocol selection exclusively relies on the minimum non-zero eigenvalue of the Laplacian matrix.
		\item[3)] For any given connected undirected graph, we present a two-step scheme to improve the anti-disturbance capability of second-order MASs. It is tractable and highly effecient when the network is unable to rearrange or expand.
	\end{itemize}
	
	The rest of this paper is organized as follows.
	In Section \ref{Sec:2}, we give some preliminaries and state the problem.
	The quantitative relations between weighted adjacency matrix, tunable gains and the \Ltwo gains, and the graph conditions for better anti-disturbance capability are given in Section \ref{Sec:3}.
	In Section \ref{Sec:4}, simulations are given to illustrate the effectiveness of theoretical results.
	We conclude our work in Section \ref{Sec:5}.
	%%%%%%%%%%%%%%%%%%%%%%%%%%%%%%%%%%%%%%%%%%%%%%%%%%%%%%%%%%%
	
	\vspace{10pt}
	\noindent\textbf{Notations}:
	Throughout this paper, $\mathbb{N}$ indicates the set
	of nonnegative integers, $\mathbb{R}$ denotes the set of real numbers, $\mathbb{R}^+$ is the set of positive real numbers, $\mathbb{R}^n$ is the
	$n$-dimensional real column vector space, 
	$\mathbb{R}^{m\times n}$ represents the $m\times n$ real matrix 
	space.
	Denote the all-one and all-zero matrices with appropriate dimensions by $\mathbf{1}$ and $\mathbf{0}$, respectively.
	Specifically, $\mathbf{1}_n$ and $\mathbf{0}_n$ refer to the $n\times1$
	all ones and all zeros column vectors, respectively.
	%$\|x\|$ is the $l_2$-norm of vector $x$.
	Let $I_n$ be the $n$-dimensional identity matrix.
	$\mathbf{j}$ stands for the imaginary unit.
	For a matrix $X$, $X^\top$ labels its transpose,
	$X^H$ denotes its conjugate transpose, and $\sigma_{max}(X)$ represents its maximum singular value.
	For a Hermitian matrix $X$, $\lambda_{max}(X)$ denotes its maximum eigenvalue.
	$X\in\mathbb{R}^{n\times n}$ is orthogonal if $X^\top X=XX^\top=I_n$.
	%For a matrix $X\in\mathbb{R}^{n \times n}$ $(\in\mathbb{C}^{n \times n})$, $\rho(X)$ is the spectral radius, $X>0$ means that $X$ is positive definite 
	%and $X\geq 0$ implies that $X$ is positive semi-definite.
	$\text{diag}\{a_1,a_2,\dots,a_n\}$ designates a diagonal 
	matrix, where $a_i$ is the $i_{\text{th}}$ diagonal element. 
	Define a set $\mathcal{I}_n=\{1,2,\dots,n\}$.
	Null set is represented by $\emptyset$.
	For the given set $\mathcal{R}_1$ and $\mathcal{R}_2$, $\mathcal{R}_1\cup\mathcal{R}_2$ and $\mathcal{R}_1\cap\mathcal{R}_2$ indicate the set union and set intersection, respectively.
	$\mathcal{L}_2\left[0,\infty\right)$ dictates the space of square-integrable vector functions, i.e., $f(t)\in\mathcal{L}_2\left[0,\infty\right)$ if and only if $\int_{0}^{\infty}f^{\top}(t)f(t)\,\mathrm{d}t<\infty$.

	%%%%%%%%%%%%%%%%%%%%%%%%%%%%%%%%%%%%%%%%%%%%%%%%%%%%%%%%%%%%%%%

	%%%%%%%%%%%%%%%%%%%%%%%%%%%%%%%%%%%%%%%%%%%%%%%%%%%%%%%%%%%%%%%
	%%%%%%%%%%%%%%%%%%%%%%%%%%%%%%%%%%%%%%%%%%%%%%%%%%%%%%%%%%%%%%%
	\section{Preliminaries and problem statement}\label{Sec:2}
	%%%%%%%%%%%%%%%%%%%%%%%%%%%%%%%%%%%%%%%%%%%%%%%%%%%%%%%%%%%%%%%
	%%%%%%%%%%%%%%%%%%%%%%%%%%%%%%%%%%%%%%%%%%%%%%%%%%%%%%%%%%%%%%%
	
	%%%%%%%%%%%%%%%%%%%%%%%%%%%%%%%%%%%%%%%%%%%%%%%%%%%%%%%%%%%%%%%
	\subsection{Preliminaries}
	%%%%%%%%%%%%%%%%%%%%%%%%%%%%%%%%%%%%%%%%%%%%%%%%%%%%%%%%%%%%%%%
	Let $\mathcal{G}=(\mathcal{V},\mathcal{E},\mathcal{A})$ be a weighted undirected graph with $n$ vertices, where $\mathcal{V}=\{s_1,s_2,\dots,s_n\}$ is the set of vertices, $\mathcal{E}\subseteq\mathcal{V}\times\mathcal{V}$ is the set of edges, and $\mathcal{A}=[a_{ij}]_{n\times n}$ is the weighted adjacency matrix with $a_{ij}=a_{ji}\geq0$.
	%An edge of $\mathcal{G}$ is denoted by
	$\varepsilon_{ij}=(s_i,s_j)\in\mathcal{E}$ if and only if there exist information exchanges between vertices $s_i$ and $s_j$.
	The adjacency element associated with the edge $\varepsilon_{ij}$ is $a_{ij}$, and $a_{ij}>0$ if and only if $\varepsilon_{ij}\in\mathcal{E}$.
	%The adjacency elements associated with the edges are positive,
	%i.e., $\varepsilon_{ij}\in\mathcal{E}$ $\Leftrightarrow$ $a_{ij}=a_{ji}>0$.
	%Since the graph $\mathcal{G}$ is weighted undirected, it means that
	%$\varepsilon_{ij}\in\mathcal{E}\Leftrightarrow\varepsilon_{ji}\in\mathcal{E}$ and the associated adjacency matrix $\mathcal{A}$ is symmetric, i.e., $\mathcal{A}=\mathcal{A}^{\top}$.
	Moreover, suppose that $\mathcal{G}$ has no self-cycles for every node, i.e., $a_{ii}=0$.
	A \textit{path} between two distinct vertices $v_i$ and $v_j$ is a finite-ordered sequence of distinct edges in $\mathcal{G}$ with the 
	form $(v_i,v_{k_1}),(v_{k_1},v_{k_2}),\dots,(v_{k_l},v_j)$.
	An undirected graph is called \textit{connected} if there exists a path between any two distinct vertices of the graph. 
	%The degree matrix $\mathcal{D}=[d_{ij}]_{n\times n}$ is a diagonal matrix with
	%$d_{ii}=\sum_{j=1}^na_{ij}$. 
	%For convenience, let $d_{max}=\max_{i\in\mathcal{I}_n}\{d_{ii}\}$.
	%$\mathcal{G}$ is called \textit{balanced} if $\sum_{i\neq j}a_{ij}=\sum_{i\neq j}a_{ji}$ for
	%all $i\in \mathcal{I}_n$.
	The Laplacian matrix $L=[l_{ij}]\in\mathbb{R}^{n\times n}$ associated with graph $\mathcal{G}$ is defined as $l_{ii}=\sum_{j=1, j\neq i}^{n}a_{ij}$ and $l_{ij}=-a_{ij}$, $j\neq i$.
	
	The following definitions and lemmas will be utilized to establish the main results.
	\begin{definition}[\cite{Silva2016}]\label{Def:0}
		For an undirected graph $\mathcal{G}$ with $n$ vertices, the network density $d$ is defined as $$d=\frac{\epsilon}{\dfrac{1}{2}n(n-1)},$$ where $\epsilon$ represents the total number of undirected edges in the graph $\mathcal{G}$ and $\frac{1}{2}n(n-1)$ is the maximum theoretical number of undirected edges between the $n$ vertices.
	\end{definition}
	
	\begin{lemma}[\cite{Jadbabaie2003}]\label{Lem:1}
		For the Laplacian matrix $L$ associated with the undirected graph $\mathcal{G}$, $L$ possesses a simple zero eigenvalue associated with eigenvector $\mathbf{1}_n$ if and only if $\mathcal{G}$ is connected.
		In addition, all the other non-zero eigenvalues are positive.
	\end{lemma}
	
	Denote the eigenvalues of Laplacian matrix $L$ associated with graph $\mathcal{G}$ by $\lambda_i$, $i\in\mathcal{I}_n$. 
	For convenience, if $\mathcal{G}$ is undirected and connected, suppose that $0=\lambda_1<\lambda_2\leq\cdots\leq\lambda_n$ and let $\Gamma=\{\lambda_2,\dots,\lambda_n\}$ be the set of all non-zero eigenvalues of $L$.
	
	%\begin{lemma}\label{Lem:3}
	%	Let $\mathcal{G}$ be a connected graph with normal Laplacian matrix 
	%	$L$, i.e.,  
	%	$L^{\top}L=LL^{\top}$. 
	%	Then $\mathcal{G}$ is weighted balanced.
	%\end{lemma}
	
	\begin{lemma}[\cite{Liu2012}]\label{Lem:2}
		Let $\mathcal{G}$ be a connected undirected graph with a Laplacian matrix $L\in\mathbb{R}^{n\times n}$ and $\Phi_n=[\varphi_{ij}]\in\mathbb{R}^{n\times n}$ be a symmetric matrix whose elements are given as 
		\begin{equation*}
			\varphi_{ij}=\begin{dcases}
				\frac{n-1}{n}, & i=j, \\
				-\frac{1}{n}, & i\neq j.
			\end{dcases}
		\end{equation*}
		%    \begin{equation}\label{eq:0.1}
			%        \Phi_n=\begin{bmatrix}
				%        	\dfrac{n-1}{n} & -\dfrac{1}{n} & \cdots & -\dfrac{1}{n} \\[2ex]
				%        	-\dfrac{1}{n} & \dfrac{n-1}{n} & \cdots & -\dfrac{1}{n} \\[0.5ex]
				%        	\vdots & \vdots & \ddots & \vdots \\[0.5ex]
				%        	-\dfrac{1}{n} & -\dfrac{1}{n} & \cdots & \dfrac{n-1}{n} \\[0.5ex]
				%        \end{bmatrix},
			%    \end{equation}
		%    then the following statements hold$:$
		%    
		%    $a)$ The eigenvalues of $\Phi_n$ are $1$ with multiplicity $n-1$ and $0$ with multiplicity $1$. 
		%    The vectors $\mathbf{1}_n^{\top}$ and $\mathbf{1}_n$ are the left and the right eigenvectors of $\Phi_n$ associated with the zero eigenvalue, respectively.
		Then, there exists an orthogonal matrix $Q\in\mathbb{R}^{n\times n}$ with the last column being $\frac{\mathbf{1}_n}{\sqrt{n}}$ such that 
		\begin{equation}\label{eq:0.2}
			Q^{\top}\Phi_nQ=\bar{\Phi}_n=\begin{bmatrix}
				I_{n-1} & \mathbf{0}_{n-1} \\
				\mathbf{0}_{n-1}^{\top} & 0
			\end{bmatrix}
		\end{equation} 
		and
		\begin{equation}\label{eq:0.3}
			Q^{\top}LQ=\bar{L}=\begin{bmatrix}
				\bar{L}_1 & \mathbf{0}_{n-1} \\
				\mathbf{0}_{n-1}^{\top} & 0
			\end{bmatrix},
		\end{equation} where $\bar{L}_1\in\mathbb{R}^{(n-1)\times (n-1)}$ is positive definite.
	\end{lemma}
	\begin{definition}[\cite{Chen2000}]\label{Def:1}
		The $H_\infty$ norm of an asymptotically stable continuous-time transfer matrix $T(s)$ is defined as
		\begin{equation*}
			\|T(s)\|_\infty= \sup_{\upsilon\in\mathbb{R}} 
			\sigma_{max}\big[T(\mathbf{j}\upsilon) \big].
		\end{equation*}
	\end{definition}
	%%%%%%%%%%%%%%%%%%%%%%%%%%%%%%%%%%%%%%%%%%%%%%%%%%%%%%%%%%%%%%%
	\subsection{Problem statement}
	%%%%%%%%%%%%%%%%%%%%%%%%%%%%%%%%%%%%%%%%%%%%%%%%%%%%%%%%%%%%%%%
	In this paper, we consider a MAS consisting of $n$ agents with double-integrator dynamics
	\begin{equation}\label{eq:1.1}
		\begin{aligned}
			\dot{x}_i(t) & =v_i(t), \\
			\dot{v}_i(t) & =u_i(t)+\omega_i(t), i\in\mathcal{I}_n,
		\end{aligned}
	\end{equation}
	where $x_i(t)\in\mathbb{R}$, $v_i(t)\in\mathbb{R}$, $u_i(t)\in\mathbb{R}$, and $\omega_i(t)\in\mathbb{R}$ are the position-like, velocity-like, control input and external disturbance of the $i_{\text{th}}$ agent, respectively. In addition, we suppose that $\omega_i(t)\in\mathcal{L}_{2}\left[0,\infty\right)$.
	% Each agent is subject to the external bounded disturbance $\omega_i(t)\in \mathcal{L}_{2}\left[0,\infty\right)$, i.e., $\int_0^\infty \omega_i^{2}(t)\,\mathrm{d}t<\infty$.
	
	The following definition of second-order consensus given in \cite{Yu2010} is useful.
	\begin{definition}[\cite{Yu2010}]\label{Def:2}
		The second-order MAS \eqref{eq:1.1} is said to reach consensus if
		\begin{equation*}
			\begin{gathered}
				\lim_{t\to \infty}\Vert x_i(t)-x_j(t)\Vert=0, \forall i,j\in\mathcal{I}_n,\\
				\lim_{t\to \infty}\Vert v_i(t)-v_j(t)\Vert=0, \forall i,j\in\mathcal{I}_n,
			\end{gathered}
		\end{equation*}
		for any initial condition.
	\end{definition}
	
	In the absence of disturbance, i.e., $\omega_i(t)=0$, as one can observe from \cite{Zhu2009} and \cite{Yu2010}, both the general absolute and relative velocity protocols \eqref{eq:1.2} and \eqref{eq:1.3} can solve the second-order consensus under the condition that $\mathcal{G}$ is a connected undirected graph.
	%In order to achieve consensus for the system \eqref{eq:1.1} when $\omega_i(t)=0$, $\forall i\in\mathcal{I}_n$, two classic consensus protocols using absolute and relative velocity information are respectively proposed in \cite{Xie2007} and \cite{Ren2007}.
	%In this paper, we use the more general protocols proposed in \cite{Zhu2009,Yu2010} taking the forms as
	\begin{equation}\label{eq:1.2}
		u_i(t)=\alpha\sum_{j=1}^n{a_{ij}\big[x_j(t)-x_i(t)\big]}-\beta v_i(t),
	\end{equation}
	\begin{equation}\label{eq:1.3}
		u_i(t)=\alpha\sum_{j=1}^n{a_{ij}\big[x_j(t)-x_i(t)\big]}+\beta \sum_{j=1}^n{a_{ij}\big[v_j(t)-v_i(t)\big]},
	\end{equation}
	where $a_{ij}$ is the $(i,j)_{\text{th}}$ entry of the weighted adjacency matrix $\mathcal{A}$ associated with undirected graph $\mathcal{G}$, and the positive constant $\alpha$ and $\beta$ are tunable gains. 
	
	Different from existing works, we restrict our attention to investigate which one of the above protocols has better anti-disturbance capability.
	%how to select between the above two protocols leading to better anti-disturbance capability.
	And we aim to bring forward simple graph conditions for protocol selection.
	%It facilitates us optimizing consensus performance by selecting existing protocols instead of designing protocols, which is handy in practice.
	%Drawing from these insights, we consider two second-order consensus protocols taking the following general forms
	
	%\begin{remark}
	%	Consider a connected undirected graph $\mathcal{G}$. As detailed in \cite{Zhu2009,Yu2010}, the velocity-like $v_i(t)\to 0, \forall i\in\mathcal{I}_n$, as $t\to \infty$ if protocol \eqref{eq:1.2} is employed.
	%%	protocol $\mathcal{P}_1$ and $\mathcal{P}_2$ both enable the MAS \eqref{eq:1.1} to achieve second-order consensus without external disturbances. In particular, 
	%	 In the other hand, we have $v_i(t)\to \frac{1}{n}\sum_{j=1}^{n}v_j(0), \forall i\in\mathcal{I}_n$, as $t\to \infty$ under protocol \eqref{eq:1.3}.
	%%	 However, we appraise a protocol only on the extent to which the disturbance corrupts consensus regardless of the consensus value.
	%\end{remark}
	%Because the consensus behavior of system \eqref{eq:1.1} is hindered by disturbances $\omega_i(t)$, we 
	Define
	\begin{equation}\label{eq:1.4}
		y_i^x(t)=x_i(t)-\frac{1}{n}\sum_{j=1}^nx_j(t),~i\in\mathcal{I}_n,
	\end{equation}
	and
	\begin{equation}\label{eq:1.5}
		y_i^v(t)=v_i(t)-\frac{1}{n}\sum_{j=1}^nv_j(t),~i\in\mathcal{I}_n,
	\end{equation}
	for each agent to measure the consensus error of position and velocity, respectively. Aggregating the outputs of all agents into a vector $y(t)\in\mathbb{R}^{2n}$ gives rise to
	\begin{equation}\label{eq:1.6}
		y(t)=\begin{bmatrix}
			y^x(t) \\ y^v(t)
		\end{bmatrix},
	\end{equation}
	where the agglomerate vector $y^x(t)=\big[y_1^x(t),\dots,y_n^x(t)\big]^{\top}$ and $y^v(t)=\big[y_1^v(t),\dots,y_n^v(t)\big]^{\top}$ denote the collective position error and collective velocity error, respectively.
	Substituting the protocols \eqref{eq:1.2} and \eqref{eq:1.3} into \eqref{eq:1.1}, respectively, and taking \eqref{eq:1.6} into consideration yield the closed-loop systems \eqref{eq:1.7} and \eqref{eq:1.8}, respectively. Here, the nominal output $y(t)$ is termed as collective consensus error.
	\begin{align}
		&\begin{dcases}\label{eq:1.7}
			\begin{bmatrix}
				\dot{x}(t) \\ \dot{v}(t)
			\end{bmatrix}=\begin{bmatrix}
				\mathbf{0} & I_{n} \\ -\alpha L & -\beta I_{n}
			\end{bmatrix}\begin{bmatrix}
				x(t) \\ v(t)
			\end{bmatrix}+\begin{bmatrix}
				\mathbf{0} \\ I_{n}
			\end{bmatrix}\omega(t), \\
			y(t)=\begin{bmatrix}
				\Phi_n & \mathbf{0} \\ \mathbf{0} & \Phi_n
			\end{bmatrix}\begin{bmatrix}
				x(t) \\ v(t)
			\end{bmatrix},
		\end{dcases}\\
		&\begin{dcases}\label{eq:1.8}
			\begin{bmatrix}
				\dot{x}(t) \\ \dot{v}(t)
			\end{bmatrix}=\begin{bmatrix}
				\mathbf{0} & I_{n} \\ -\alpha L & -\beta L
			\end{bmatrix}\begin{bmatrix}
				x(t) \\ v(t)
			\end{bmatrix}+\begin{bmatrix}
				\mathbf{0} \\ I_{n}
			\end{bmatrix}\omega(t), \\
			y(t)=\begin{bmatrix}
				\Phi_n & \mathbf{0} \\ \mathbf{0} & \Phi_n
			\end{bmatrix}\begin{bmatrix}
				x(t) \\ v(t)
			\end{bmatrix},
		\end{dcases}
	\end{align}
	where $x(t)=\big[x_1(t),\dots,x_n(t)\big]^{\top}$, $v(t)=\big[v_1(t),\dots,v_n(t)\big]^{\top}$, $\omega(t)=\big[\omega_1(t),\dots,\omega_n(t)\big]^{\top}$, and $L$ is the Laplacian matrix associated with graph $\mathcal{G}$.
	
	In this paper, we respectively use the $\mathcal{L}_2$ gains from the  disturbance $\omega(t)$ to collective consensus error $y(t)$ of systems \eqref{eq:1.7} and \eqref{eq:1.8} to measure the anti-disturbance capabilities of the MASs \eqref{eq:1.1}-\eqref{eq:1.2} and \eqref{eq:1.1}-\eqref{eq:1.3}.
	As shown in \cite{Pirani2021}, the $\mathcal{L}_2$ gains of systems \eqref{eq:1.7} and \eqref{eq:1.8} are defined by
	\begin{equation*}
		\sup_{\substack{\omega(t)\neq0\\ \omega(t)\in \mathcal{L}_2\left[0,\infty\right)}}
			\sqrt{\frac{\int_{0}^{\infty}y^{\top}(t)y(t)\,\mathrm{d}t}{\int_{0}^{\infty}\omega^{\top}(t)\omega(t)\,\mathrm{d}t}}.
	\end{equation*}
	However, it is difficult to directly use $\mathcal{L}_2$ gain to analyze the anti-disturbance capability of the considered MASs.
	Let $T_1(s)$ and $T_2(s)$ be the transfer matrices from disturbance $\omega(t)$ to collective consensus error $y(t)$ of the systems \eqref{eq:1.7} and \eqref{eq:1.8}, respectively. 
	%Then the $\mathcal{L}_2$ gains of \eqref{eq:1.7} and \eqref{eq:1.8} are respectively equivalent to $\|T_1(s)\|_\infty$ and $\|T_2(s)\|_\infty$ which refer to the \Hinf norm of $T_1(s)$ and $T_2(s)$ \cite{Zhou1998}.
	It follows from \cite{Chen2000} that the $\mathcal{L}_2$ gains of systems \eqref{eq:1.7} and \eqref{eq:1.8} are equal to $\|T_1(s)\|_\infty$ and $\|T_2(s)\|_\infty$, respectively, where $\|T_i(s)\|_\infty$ refers to the \Hinf norm of $T_i(s)$ $(i=1,2)$. Clearly, smaller \Hinf norm of transfer matrix means better anti-disturbance capability.
	%Obviously, smaller \Hinf norm of transfer matrix implies better anti-disturbance capability.
	%\begin{remark}
	%	For a connected undirected graph $\mathcal{G}$, according to Lemma \ref{Lem:1}, it is easy to verify that system \eqref{eq:1.7} and \eqref{eq:1.8} are marginally stable .
	%    However, $\|T_1(s)\|_\infty$ and $\|T_2(s)\|_\infty$ are still well-defined because the marginally stable modals of system \eqref{eq:1.7} and \eqref{eq:1.8} are unobservable from $y(t)$ \cite{Siami2015}.
	%\end{remark}
	
	%For above configuration, we aim to address the closed-form expressions of $\|T_1(s)\|_\infty$ and $\|T_2(s)\|_\infty$ which serve the purpose of choosing a better protocol for second-order consensus. Naturally, we introduce the following definition in which the superiority of a protocol is reduced to a quantitative description.
	Note that $\|T_1(s)\|_\infty$ and $\|T_2(s)\|_\infty$ are easy to tackle. In the following, for connected undirected graph $\mathcal{G}$, we directly use $\|T_1(s)\|_\infty$ and $\|T_2(s)\|_\infty$ to characterize the anti-disturbance capabilities of MASs \eqref{eq:1.1}-\eqref{eq:1.2} and \eqref{eq:1.1}-\eqref{eq:1.3}, respectively.
	For convenience, we introduce the following definitions.
	%the \Hinf control problem of the system $\mathcal{S}_1$ and $\mathcal{S}_2$ can be well solved by designing proper tunable gain $\alpha$ and $\beta$ through \Hinf control theory and LMIs technique. However, our study goes into a quite different direction. 
	
	%We restrict attention here to 
	
	\begin{definition}\label{Def:3}
		For the second-order MAS \eqref{eq:1.1}, assume that $G$ is a connected undirected graph.
		\begin{itemize}
			\item[1)] The protocol \eqref{eq:1.2} is said to outperform the protocol \eqref{eq:1.3} if the anti-disturbance capability of MAS \eqref{eq:1.1}-\eqref{eq:1.2} outperforms that of MAS \eqref{eq:1.1}-\eqref{eq:1.3}, i.e., $\|T_1(s)\|_\infty\leq \|T_2(s)\|_\infty$$;$
			\item[2)] The protocol \eqref{eq:1.3} is said to outperform the protocol \eqref{eq:1.2} if the anti-disturbance capability of MAS \eqref{eq:1.1}-\eqref{eq:1.3} outperforms that of MAS \eqref{eq:1.1}-\eqref{eq:1.2}, i.e., $\|T_2(s)\|_\infty\leq \|T_1(s)\|_\infty$$;$
			\item[3)] The protocol \eqref{eq:1.2} is said to perform as well as the protocol \eqref{eq:1.3} if the MASs \eqref{eq:1.1}-\eqref{eq:1.3} and \eqref{eq:1.1}-\eqref{eq:1.2} have the same anti-disturbance capability, i.e., $\|T_2(s)\|_\infty\equiv \|T_1(s)\|_\infty$$.$
		\end{itemize}
		%	For two given protocols $\mathcal{M}$ and $\mathcal{N}$, $\mathcal{M}\succ\mathcal{N}$ denotes that protocol $\mathcal{M}$ outperforms protocol $\mathcal{N}$, and $\mathcal{M}\sim\mathcal{N}$ represents that protocol $\mathcal{M}$ performs as well as protocol $\mathcal{N}$. For the second-order MAS \eqref{eq:1.1} on a connected undirected graph $\mathcal{G}$, we state that
		%	
		%	$1)$ \eqref{eq:1.2} $\succ$ \eqref{eq:1.3} if $\|T_1(s)\|_\infty\leq\|T_2(s)\|_\infty, \forall\alpha,\beta>0$$;$
		%	
		%	$2)$ \eqref{eq:1.3} $\succ$ \eqref{eq:1.2} if $\|T_2(s)\|_\infty\leq\|T_1(s)\|_\infty, \forall\alpha,\beta>0$$;$
		%	
		%	$3)$ \eqref{eq:1.2} $\sim$ \eqref{eq:1.3} if $\|T_1(s)\|_\infty\equiv\|T_2(s)\|_\infty, \forall\alpha,\beta>0$$.$
		%	\begin{equation*}
			%	    \begin{dcases}
				%			\mathcal{P}_1~\textit{outperforms}~\mathcal{P}_2 & \textit{if}~ \rho_{\mathcal{S}_1}\leq\rho_{\mathcal{S}_2}, \\
				%            \mathcal{P}_2~\textit{outperforms}~\mathcal{P}_1 & \textit{if}~ \rho_{\mathcal{S}_1}\geq\rho_{\mathcal{S}_2}, \\
				%            \mathcal{P}_1~\textit{performs as well as}~\mathcal{P}_2 & \textit{if}~ \rho_{\mathcal{S}_1}\equiv\rho_{\mathcal{S}_2}.
				%		\end{dcases}
			%	\end{equation*}
	\end{definition}
	%In a word, we prefer to select a protocol that makes the MAS \eqref{eq:1.1} more robust.
	%\begin{definition}[Second-order consensus]
	%	
	%\end{definition}

	%%%%%%%%%%%%%%%%%%%%%%%%%%%%%%%%%%%%%%%%%%%%%%%%%%%%%%%%%%%%%%%%%%%%%%
	%%%%%%%%%%%%%%%%%%%%%%%%%%%%%%%%%%%%%%%%%%%%%%%%%%%%%%%%%%%%%%%%%%%%%%
	\section{Main results}\label{Sec:3}
	%%%%%%%%%%%%%%%%%%%%%%%%%%%%%%%%%%%%%%%%%%%%%%%%%%%%%%%%%%%%%%%%%%%%%%
	%%%%%%%%%%%%%%%%%%%%%%%%%%%%%%%%%%%%%%%%%%%%%%%%%%%%%%%%%%%%%%%%%%%%%%
	In this section, we will establish the graph conditions to tell which protocol has better anti-disturbance for MAS \eqref{eq:1.1}.
	%intend to establish the closed-form expressions of $\|T_1(s)\|_\infty$ and $\|T_2(s)\|_\infty$ for the system \eqref{eq:1.7} and \eqref{eq:1.8}, respectively, so as to quantitatively compare the anti-disturbance capabilities of MASs \eqref{eq:1.1}-\eqref{eq:1.2} and \eqref{eq:1.1}-\eqref{eq:1.3}. And on this basis, we want to 
	%we measure the anti-disturbance capability of MASs by its \Hinf performance.
	%Then we respectively ascertain the closed-form expression of \Hinf performance for second-order continuous-time MAS \eqref{eq:1.1} with two different protocols \eqref{eq:1.2} and \eqref{eq:1.3}. Based on this formulation, we bring forward a simple graph condition pertaining to protocol selection for better \Hinf performance.
	Prior to stating the main results, we propose a useful lemma.
	\begin{lemma}\label{Lem:3}
		Let $g_1(t)=\dfrac{1}{(\alpha t)^2}$, $g_2(t)=\dfrac{1}{(\alpha t)^2-(\sqrt{\Delta}-1)^2}$, and $g_3(t)=\dfrac{1}{(\alpha t)^2-(\sqrt{\Theta}-1)^2}$,
		where $\Delta=(\alpha t+1)^2-\beta^2$, $\Theta=(\alpha t+1)^2-(\beta t)^2$, and the constants $\alpha$ and $\beta$ are positive.
		Then the following statements hold$:$
		\begin{itemize}
			\item[1)] $g_1(t)$ is decreasing on $(0,+\infty)$$;$
			\item[2)] $g_2(t)$ is decreasing on $(\frac{\beta-1}{\alpha},+\infty)$$;$
			\item[3)] $g_3(t)$ is decreasing on $(0,+\infty)$ if $\beta\leq\alpha$$;$
			\item[4)] $g_3(t)$ is decreasing on $(0,\frac{2\alpha}{\beta^2-\alpha^2})$ if $\beta>\alpha$.
		\end{itemize}
	\end{lemma}
	\begin{Proof}
		Clearly, $g_1(t)$ is decreasing with $t$ on $(0,+\infty)$. Let $h_1(t)=(\alpha t)^2-(\sqrt{\Delta}-1)^2$, then it is inferred from $t\in(\frac{\beta-1}{\alpha},+\infty)$ that
		\begin{equation*}
			\begin{aligned}
				\frac{\mathrm{d}h_1(t)}{\mathrm{d}t}&=\frac{2\alpha^2t+2\alpha}{\sqrt{(\alpha t+1)^2-\beta^2}}-2\alpha \\
				&=2\alpha\sqrt{\frac{(\alpha t+1)^2}{(\alpha t+1)^2-\beta^2}}-2\alpha>0.
			\end{aligned}
		\end{equation*}
		It follows from
		\begin{equation*}
			\frac{\mathrm{d}g_2(t)}{\mathrm{d}t}=-\frac{\frac{\mathrm{d}h_1(t)}{\mathrm{d}t}}{h_1^2(t)}<0
		\end{equation*}
		that $g_2(t)$ is a decreasing function of $t$ on $(\frac{\beta-1}{\alpha},+\infty)$.
		
		Likewise, let $h_2(t)=(\alpha t)^2-(\sqrt{\Theta}-1)^2$, then we have
		\begin{equation*}
			\frac{\mathrm{d}h_2(t)}{\mathrm{d}t}=\frac{2\alpha(\alpha t+1)-2\beta^2t}{\sqrt{(\alpha t+1)^2-(\beta t)^2}}+2\beta^2t-2\alpha.
		\end{equation*}
		The second derivative of $h_2(t)$ turns out to be
		\begin{equation*}
			\begin{aligned}
				\frac{\mathrm{d}^2h_2(t)}{\mathrm{d}t^2}&=\frac{(2\alpha^2-2\beta^2)\sqrt{\Theta}-\dfrac{2\big[\alpha(\alpha t+1)-\beta^2t\big]^2}{\sqrt{\Theta}}}{\Theta}+2\beta^2 \\
				&=\frac{(2\alpha^2-2\beta^2)\Theta-2\big[\alpha(\alpha t+1)-\beta^2t\big]^2}{\Theta^{\frac{3}{2}}}+2\beta^2 \\
				%			&=\frac{-2\beta^2[(\alpha t+1)-\alpha t]^2}{\Theta^{\frac{3}{2}}}+2\beta^2 \\
				&=\frac{-2\beta^2}{\Theta^{\frac{3}{2}}}+2\beta^2.
			\end{aligned}
		\end{equation*}
		
		If $\beta\leq\alpha$, we get $\Theta>1, \forall t\in(0,+\infty)$ which means that $\frac{\mathrm{d}^2h_2(t)}{\mathrm{d}t^2}>0, \forall t\in(0,+\infty)$. Combining with the fact $\frac{\mathrm{d}h_2(t)}{\mathrm{d}t}\big\vert_{t=0}=0$ gives rise to $\frac{\mathrm{d}h_2(t)}{\mathrm{d}t}>0, \forall t\in(0,+\infty)$.
		Then
		\begin{equation}\label{eq:diff_g3}
			\frac{\mathrm{d}g_3(t)}{\mathrm{d}t}=-\frac{\frac{\mathrm{d}h_2(t)}{\mathrm{d}t}}{h_2^2(t)}<0
		\end{equation}
		holds for any $t\in(0,+\infty)$. Thus, $g_3(t)$ is decreasing on $(0,+\infty)$.
		
		Provided that $\beta>\alpha$, we have $\Theta>1, \forall t\in(0,\frac{2\alpha}{\beta^2-\alpha^2})$ which implies that $\frac{\mathrm{d}^2h_2(t)}{\mathrm{d}t^2}>0, \forall t\in(0,\frac{2\alpha}{\beta^2-\alpha^2})$. Combining with $\frac{\mathrm{d}h_2(t)}{\mathrm{d}t}\big\vert_{t=0}=0$ derives that $\frac{\mathrm{d}h_2(t)}{\mathrm{d}t}>0, \forall t\in(0,\frac{2\alpha}{\beta^2-\alpha^2})$.
		Then \eqref{eq:diff_g3} holds for any $t\in(0,\frac{2\alpha}{\beta^2-\alpha^2})$. Therefore, $g_3(t)$ is decreasing on $(0,\frac{2\alpha}{\beta^2-\alpha^2})$.\hfill $\square$
		%	As long as $\Theta>1$, the facts $\frac{\mathrm{d}^2h_2(t)}{\mathrm{d}t^2}>0$ and $\frac{\mathrm{d}h_2(t)}{\mathrm{d}t}\big\vert_{t=0}=0$
		%%	\begin{equation*}
			%%		\begin{dcases}
				%%			\frac{\mathrm{d}^2h_2(t)}{\mathrm{d}t^2}>0 \\ \frac{\mathrm{d}h_2(t)}{\mathrm{d}t}\Big\vert_{t=0}=0
				%%		\end{dcases}
			%%	\end{equation*}
		%	implies that $\frac{\mathrm{d}h_2(t)}{\mathrm{d}t}>0$. Therefore, we have

	\end{Proof}
	\subsection{Anti-disturbance capability of the second-order MAS using absolute velocity information}
	Theorem \ref{theorem:1} gives the analytic expression of the anti-disturbance capability of second-order MAS \eqref{eq:1.1} with absolute velocity protocol \eqref{eq:1.2}. 
	%Specifically, we build the relation among $\|T_1(s)\|_\infty$, the smallest non-zero eigenvalue of Laplacian matrix and the tunable gains.
	\begin{theorem}\label{theorem:1}
		%	Cosider the MAS \eqref{eq:1.1}-\eqref{eq:1.2} on a connected undirected graph $\mathcal{G}$ with the Laplacian matrix $L$. We can obtain that
		Consider MAS \eqref{eq:1.1}-\eqref{eq:1.2} in which $\mathcal{G}$ is a connected undirected graph with Laplacian matrix $L$. Then, we obtain
		\begin{equation}\label{eq:Th1}
			\|T_1(s)\|_\infty=\begin{dcases}
				\frac{1}{\sqrt{(\alpha\lambda_2)^2-\Big[\sqrt{(\alpha\lambda_2+1)^2-\beta^2}-1\Big]^2}}, & \text{if}~~0<\beta<\sqrt{(\alpha\lambda_{2})^2+2\alpha\lambda_{2}}, \\
				\frac{1}{\alpha\lambda_2}, & \text{if}~~\beta\geq\sqrt{(\alpha\lambda_{2})^2+2\alpha\lambda_{2}}.
			\end{dcases}
		\end{equation}
		%    and assume the form
		%    $0=Re(\lambda_1)< Re(\lambda_2)\leq \cdots \leq Re(\lambda_n).$
	\end{theorem}
	\begin{Proof}
		%	Although system \eqref{eq:1.3}-\eqref{eq:1.8} is marginally stable, its \Hinf norm is well-defined as long as the marginally stable subsystem is unobservable.
		%	Therefore, we need to transform the marginally stable system \eqref{eq:1.3}-\eqref{eq:1.8} to obtain its asymptotically stable subsystem and marginally stable subsystem.
		Since the undirected graph $\mathcal{G}$ is connected, it follows from Lemma \ref{Lem:2} that there exists an orthogonal matrix $Q\in\mathbb{R}^{n\times n}$ such that \eqref{eq:0.2} and \eqref{eq:0.3} hold.
		%\begin{equation}\label{eq:2.0}
		%	Q^{\top}LQ=\bar{L}=\begin{bmatrix}
			%		\bar{L}_1 & \mathbf{0}_{n-1} \\
			%		\mathbf{0}_{n-1}^{\top} & 0
			%	\end{bmatrix},
		%\end{equation}
		%	where $\bar{L}_1\in\mathbb{R}^{(n-1)\times (n-1)}$ is positive definite and has the same non-zero eigenvalues as $L$.
		%	$$U^{\top}\Phi_nU=\bar{\Phi}_n=\begin{bmatrix}
			%		I_{n-1} & \mathbf{0}_{n-1} \\
			%		\mathbf{0}_{n-1}^{\top} & 0
			%	\end{bmatrix}$$
		%    and
		%    $$U^{\top}LU=\bar{L}=\begin{bmatrix}
			%    	\bar{L}_1 & \mathbf{0}_{n-1} \\
			%    	\mathbf{0}_{n-1}^{\top} & 0
			%    \end{bmatrix},$$
		%	According to $L\mathbf{1}_{n}=\mathbf{0}_{n}$, $\Phi_n\mathbf{1}_{n}=\mathbf{0}_{n}$, and \eqref{eq:1.6}, we can obtain $Le(t)=LX(t)$ and $\Phi_ne(t)=\Phi_nX(t)$.
		%	Then combining system \eqref{eq:1.3}-\eqref{eq:1.8} with \eqref{eq:1.6} yields the consensus error system
		%	\begin{equation}\label{eq:2.2}
			%		\begin{dcases}
				%			\dot{e}(t)=-\Phi_nLe(t)+\Phi_n\Omega(t), \\
				%			y(t)=\begin{bmatrix}
					%				\Phi_n \\ -L
					%			\end{bmatrix}e(t).
				%		\end{dcases}
			%	\end{equation}
		Then introducing the following orthogonal transformation
		\begin{equation}\label{eq:2.2}
			\begin{aligned}
				\begin{bmatrix}
					\hat{x}(t) \\ \hat{v}(t)
				\end{bmatrix}&=\begin{bmatrix}
					Q^{\top} & \mathbf{0} \\ \mathbf{0} & Q^{\top}
				\end{bmatrix}\begin{bmatrix}
					x(t) \\ v(t)
				\end{bmatrix}, \\
				\hat{y}(t)&=\begin{bmatrix}
					Q^{\top} & \mathbf{0} \\ \mathbf{0} & Q^{\top}
				\end{bmatrix}y(t),\\
				\hat{\omega}(t)&=Q^{\top}\omega(t),
			\end{aligned}
		\end{equation}
		for system \eqref{eq:1.7} gives rise to
		\begin{equation}\label{eq:2.1}
			\begin{dcases}
				\begin{bmatrix}
					\dot{\hat{x}}(t) \\ \dot{\hat{v}}(t)
				\end{bmatrix}=\begin{bmatrix}
					\mathbf{0} & I_{n} \\ -\alpha \bar{L} & -\beta I_{n}
				\end{bmatrix}\begin{bmatrix}
					\hat{x}(t) \\ \hat{v}(t)
				\end{bmatrix}+\begin{bmatrix}
					\mathbf{0} \\ I_{n}
				\end{bmatrix}\hat{\omega}(t), \\
				\hat{y}(t)=\begin{bmatrix}
					\bar{\Phi}_n & \mathbf{0} \\ \mathbf{0} & \bar{\Phi}_n
				\end{bmatrix}\begin{bmatrix}
					\hat{x}(t) \\ \hat{v}(t)
				\end{bmatrix},
			\end{dcases}
		\end{equation}
		where $$\bar{L}=\begin{bmatrix}
			\bar{L}_1 & \mathbf{0}_{n-1} \\
			\mathbf{0}_{n-1}^{\top} & 0
		\end{bmatrix}.$$
		As evidenced from Lemma \ref{Lem:1} and \eqref{eq:0.3}, positive definite matrix $\bar{L}_1$ possesses the same non-zero eigenvalues of $L$ which implies that $-\bar{L}_1$ is Hurwitz stable.
		Clearly, system \eqref{eq:2.1} is composed of an asymptotically stable subsystem of order $2n-2$ and a marginally stable subsystem of order $2$.	
		Since the asymptotically stable subsystem is observable and the marginally stable subsystem is unobservable, according to \cite{Siami2015}, $\|T_1(s)\|_\infty$ is still well-difined and completely determined by the asymptotically stable subsystem. 
		Consider the asymptotically stable subsystem of \eqref{eq:2.1} taking the form of
		%	Therefore, we only need to analyze the asymptotically stable subsystem of \eqref{eq:2.1} which is as follows
		\begin{equation}\label{eq:2.5}
			\begin{dcases}
				\begin{bmatrix}
					\dot{\hat{x}}^1(t) \\ \dot{\hat{v}}^1(t)
				\end{bmatrix}=\begin{bmatrix}
					\mathbf{0} & I_{n-1}  \\ -\alpha\bar{L}_1 & -\beta I_{n-1}
				\end{bmatrix}
				\begin{bmatrix}
					\hat{x}^1(t) \\ \hat{v}^1(t)
				\end{bmatrix}+\begin{bmatrix}
					\mathbf{0} \\ I_{n-1}
				\end{bmatrix}\hat{\omega}^1(t), \\
				\hat{y}^1(t)=\begin{bmatrix}
					I_{n-1} & \mathbf{0}\\ \mathbf{0} &  I_{n-1}
				\end{bmatrix}\begin{bmatrix}
					\hat{x}^1(t) \\ \hat{v}^1(t)
				\end{bmatrix},
			\end{dcases}
		\end{equation}
		where $\hat{x}^1(t)=\begin{bmatrix}I_{n-1} & \mathbf{0}_{n-1}\end{bmatrix}\hat{x}(t)$, $\hat{v}^1(t)=\begin{bmatrix}I_{n-1} & \mathbf{0}_{n-1}\end{bmatrix}$\ \ \ $\hat{v}(t)$ and $\hat{\omega}^1(t)=\begin{bmatrix}I_{n-1} & \mathbf{0}_{n-1}\end{bmatrix}\hat{\omega}(t).$
		Since $\bar{L}_1$ is positive definite, there exists an orthogonal matrix 
		$V\in\mathbb{R}^{(n-1)\times (n-1)}$ such that $V^\top\bar{L}_1V=\varLambda$, where $\varLambda=\text{diag}\{\lambda_2,\lambda_3,\dots,\lambda_n\}$ is composed of 
		the non-zero eigenvalues of $L$. Then performing the following orthogonal transformation
		\begin{equation}\label{eq:2.7}
			\begin{aligned}
				\begin{bmatrix}
					\tilde{x}(t) \\ \tilde{v}(t)
				\end{bmatrix}&=\begin{bmatrix}
					V^{\top} & \mathbf{0} \\ \mathbf{0} & V^{\top}
				\end{bmatrix}\begin{bmatrix}
					\hat{x}^1(t) \\ \hat{v}^1(t)
				\end{bmatrix}, \\
				\tilde{y}(t)&=\begin{bmatrix}
					V^{\top} & \mathbf{0} \\ \mathbf{0} & V^{\top}
				\end{bmatrix}\hat{y}^1(t),\\
				\tilde{\omega}(t)&=V^{\top}\hat{\omega}^1(t), 
			\end{aligned}
		\end{equation}
		for the system \eqref{eq:2.5} provides
		\begin{equation}\label{eq:2.8}
			\begin{dcases}
				\begin{bmatrix}
					\dot{\tilde{x}}(t) \\ \dot{\tilde{v}}(t)
				\end{bmatrix}=\begin{bmatrix}
					\mathbf{0} & I_{n-1}  \\ -\alpha\varLambda & -\beta I_{n-1}
				\end{bmatrix}
				\begin{bmatrix}
					\tilde{x}(t) \\ \tilde{v}(t)
				\end{bmatrix}+\begin{bmatrix}
					\mathbf{0} \\ I_{n-1}
				\end{bmatrix}\tilde{\omega}(t), \\
				\tilde{y}(t)=\begin{bmatrix}
					I_{n-1} & \mathbf{0}\\ \mathbf{0} &  I_{n-1}
				\end{bmatrix}\begin{bmatrix}
					\tilde{x}(t) \\ \tilde{v}(t)
				\end{bmatrix}.
			\end{dcases}
		\end{equation}
		Denote the transfer matrix of the system \eqref{eq:2.1}, \eqref{eq:2.5} and \eqref{eq:2.8} by $T_3(s)$, $T_4(s)$ and $T_5(s)$, respectively.
		Through simple calculation, it is easy to verify that
		\begin{equation}\label{eq:T1=T5}
			\|T_1(s)\|_\infty=\|T_3(s)\|_\infty=\|T_4(s)\|_\infty=\|T_5(s)\|_\infty.
		\end{equation}
		Then we turn to compute $\|T_5(s)\|_\infty$.
		
		The transfer matrix $T_5(s)$ is shown as
		\begin{equation*}
			T_5(s)=\left[\begin{array}{ccc}
				\Xi_2& &\\
				& \ddots &\\
				& & \Xi_n \\
				\hdashline
				\Upsilon_2& &\\
				& \ddots &\\
				& & \Upsilon_n
			\end{array}\right],
		\end{equation*}
		where $\Xi_i=\dfrac{1}{s^2+\beta s+\alpha \lambda_i}$ and $\Upsilon_i=\dfrac{s}{s^2+\beta s+\alpha \lambda_i}$, $i=2,\dots,n$.
		It can be readily obtained that
		\begin{equation*}
			T_5^H(\mathbf{j}\upsilon)T_5(\mathbf{j}\upsilon)=\begin{bmatrix}
				\delta_2(\upsilon)& &\\
				& \ddots &\\
				& & \delta_n(\upsilon)
			\end{bmatrix}
		\end{equation*}
		is diagonal, where $$\delta_i(\upsilon)=\frac{1+\upsilon^2}{(\alpha\lambda_i-\upsilon^2)^2+(\beta \upsilon)^2}>0,~i=2,\dots,n,~\upsilon\in\mathbb{R}.$$
		According to Definition \ref{Def:1} and \eqref{eq:T1=T5}, $\|T_1(s)\|_\infty$ can be treated as
		\begin{equation}
			\begin{split}\label{eq:A-1}
				\|T_1(s)\|_\infty&=\|T_5(s)\|_\infty \\ 
				&=\sup_{\upsilon\in\mathbb{R}} 
				\sqrt{\lambda_{max}\big[T_5^H(\mathbf{j}\upsilon)T_5(\mathbf{j}\upsilon)\big]} \\
				&=\max_{i=2,\dots,n}\sqrt{\sup_{\upsilon\in\mathbb{R}}\delta_{i}(\upsilon)}.
			\end{split}
		\end{equation}
		It is inferred from 
		\begin{equation}\label{eq:diff_delta_i}
			\frac{\mathrm{d}\delta_i(\upsilon)}{\mathrm{d}\upsilon}=\frac{-2\upsilon\big[\upsilon^4+2\upsilon^2+\beta^2-(\alpha\lambda_{i})^2-2\alpha\lambda_{i}\big]}{\big[\upsilon^4+(\beta^2-2\alpha\lambda_{i})\upsilon^2+(\alpha\lambda_{i})^2\big]^2}=0
		\end{equation}
		that
		\begin{equation}\label{eq:numerator_diff_delta}
			\upsilon\big[\upsilon^4+2\upsilon^2+\beta^2-(\alpha\lambda_{i})^2-2\alpha\lambda_{i}\big]=0.
		\end{equation}
		Obviously, $\sup_{\upsilon\in\mathbb{R}}\delta_{i}(\upsilon)$ depends on $\beta^2-(\alpha\lambda_{i})^2-2\alpha\lambda_{i}$.
		If $\beta\geq\sqrt{(\alpha\lambda_{i})^2+2\alpha\lambda_{i}}$, by solving \eqref{eq:numerator_diff_delta}, it is easy to obtain
		\begin{equation*}%\label{eq:delta1}
			\sup_{\upsilon\in\mathbb{R}}\delta_{i}(\upsilon)=\delta_{i}(\upsilon^*_{i,1})=g_1(\lambda_i)=\frac{1}{(\alpha \lambda_{i})^2},
		\end{equation*}
		where $\upsilon^*_{i,1}=0$.
		If $\beta<\sqrt{(\alpha\lambda_{i})^2+2\alpha\lambda_{i}}$, by solving \eqref{eq:numerator_diff_delta}, it can be readily concluded that
		\begin{equation*}%\label{eq:delta2}
			\begin{split}
				\sup_{\upsilon\in\mathbb{R}}\delta_{i}(\upsilon)=&\delta_{i}(\upsilon^*_{i,2})=\delta_{i}(\upsilon^*_{i,3})=g_2(\lambda_i) \\
				=&\frac{1}{(\alpha\lambda_i)^2-(\sqrt{\Delta_i}-1)^2}
			\end{split}
		\end{equation*}
		where $\upsilon^*_{i,2},\upsilon^*_{i,3}=\pm\sqrt{\sqrt{\Delta_i}-1}$, and $\Delta_i=(\alpha\lambda_i+1)^2-\beta^2$.
		The trajectories of the function $\delta_{i}(\upsilon)$ under the conditions $\beta\geq\sqrt{(\alpha\lambda_{i})^2+2\alpha\lambda_{i}}$ and $\beta<\sqrt{(\alpha\lambda_{i})^2+2\alpha\lambda_{i}}$ are displayed in Fig. \ref{fig1}.

		\begin{figure}[!ht]
			\centering
			\hspace{-1cm}
			\includegraphics[width=0.48\textwidth,height=0.36\textwidth]{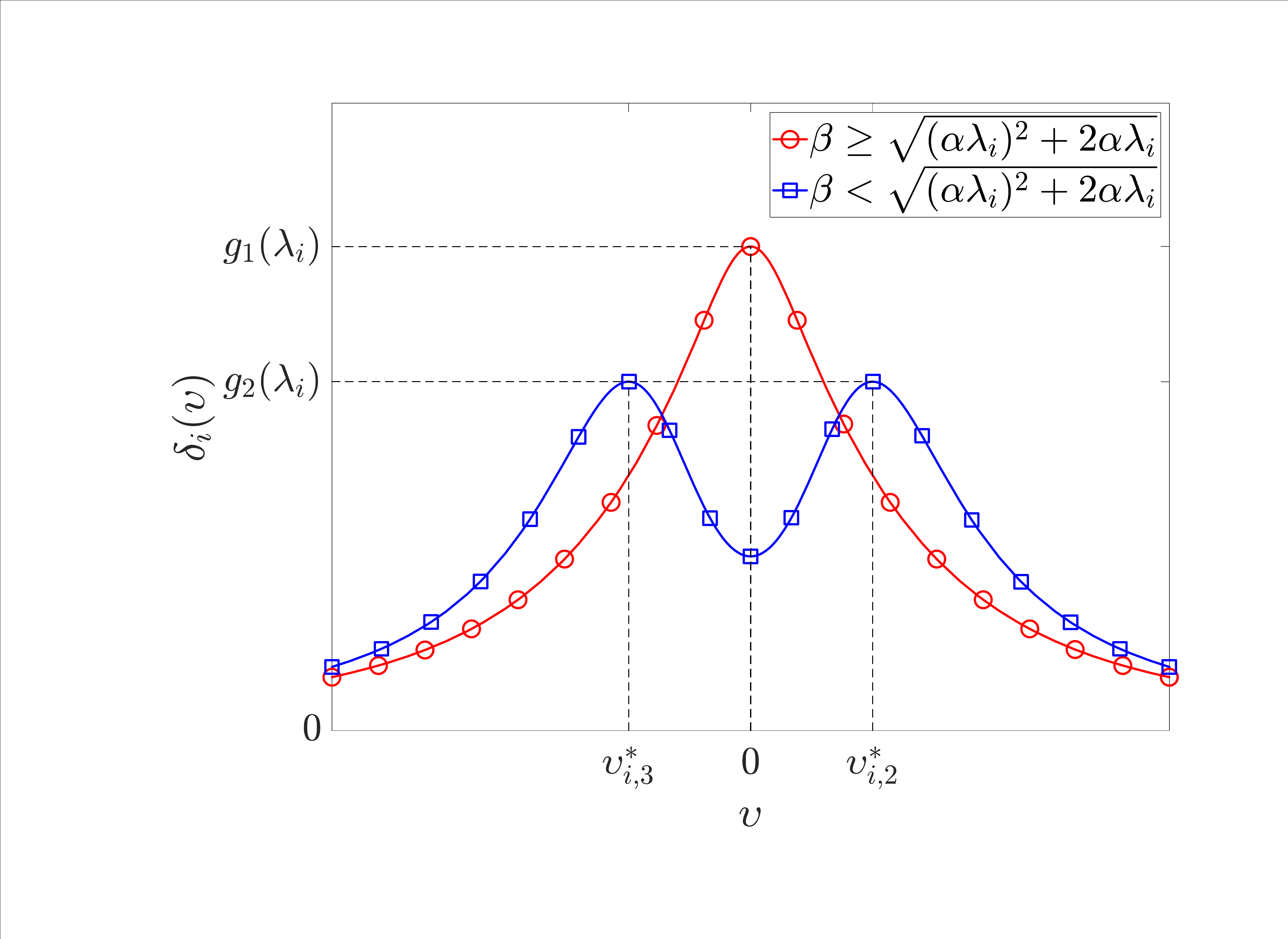}
			\caption{Trajectories of $\delta_{i}(\upsilon)$ under different conditions.\label{fig1}}
		\end{figure}
		
		Building on these preliminary observations, we refer to
		\begin{equation*}
			\mathcal{R}_1=\Big\{r\in\Gamma~\big|~\beta\geq\sqrt{(\alpha r)^2+2\alpha r}\Big\}
		\end{equation*}
		and
		\begin{equation*}
			\mathcal{R}_2=\Big\{r\in\Gamma~\big|~\beta<\sqrt{(\alpha r)^2+2\alpha r}\Big\} 
		\end{equation*}
		as two sets of non-zero eigenvalues of $L$.
		Obviously, $\mathcal{R}_1\cup\mathcal{R}_2=\Gamma$ and  
		$\mathcal{R}_1\cap\mathcal{R}_2=\emptyset$ hold.
		%    Furthermore, it is easy to verify that
		%    \begin{equation}\label{eq:R2}
			%      r>\frac{\sqrt{1+\beta^2}-1}{\alpha}>\frac{\beta-1}{\alpha}, \forall r\in\mathcal{R}_2.
			%    \end{equation}
		%    Furthermore, we can conclude that
		%    \begin{equation}\label{eq:case1}
			%    	\sup_{\upsilon\in\mathbb{R}}\delta_{i}(\upsilon)=g_1{(\lambda_{i})}, \forall \lambda_{i}\in\mathcal{R}_1,
			%    \end{equation}
		%    and
		%    \begin{equation}\label{eq:case2}
			%    	\sup_{\upsilon\in\mathbb{R}}\delta_{i}(\upsilon)=g_2{(\lambda_{i})}, \forall \lambda_{i}\in\mathcal{R}_2.
			%    \end{equation}
		We will complete the proof by enumeration.
		
		(\uppercase\expandafter{\romannumeral1})
		$\mathcal{R}_1=\Gamma$ and $\mathcal{R}_2=\emptyset$;
		
		In this case, we have $\beta\geq\sqrt{(\alpha \lambda_{i})^2+2\alpha\lambda_{i}}, \forall i\in\{2,\dots,n\}$ which leads to $\sup_{\upsilon\in\mathbb{R}}\delta_{i}(\upsilon)=g_1{(\lambda_{i})}, \forall i\in\{2,\dots,n\}$.
		Because $0<\lambda_2\leq\cdots\leq\lambda_n$ and $g_1{(t)}$ is decreasing on $(0,+\infty)$, one can deduce that $\max_{i=2,\dots,n}g_1{(\lambda_{i})}=g_1(\lambda_2)$.
		Therefore, \eqref{eq:A-1} can be written as
		%	\begin{equation}\label{eq:case1}
			%		\sup_{\upsilon\in\mathbb{R}}\delta_{i}(\upsilon)=g_1{(\lambda_{i})}, \forall i\in\{2,\dots,n\}.
			%	\end{equation}
		\begin{equation*}
			%		\begin{aligned}
				\|T_1(s)\|_\infty=\max_{i=2,\dots,n}\sqrt{g_1{(\lambda_{i})}}=\sqrt{g_1(\lambda_2)}=\frac{1}{\alpha\lambda_2}.
				%		\end{aligned}
		\end{equation*}
		
		(\uppercase\expandafter{\romannumeral2})
		$\mathcal{R}_1=\emptyset$ and $\mathcal{R}_2=\Gamma$;
		
		%    \begin{equation}\label{eq:case2}
			%    	\sup_{\upsilon\in\mathbb{R}}\delta_{i}(\upsilon)=g_2{(\lambda_{i})}, \forall i\in\{2,\dots,n\}.
			%    \end{equation}
		%    Moreover, it follows from $\mathcal{R}_2=\Gamma$ that $$r>\frac{\sqrt{1+\beta^2}-1}{\alpha}>\frac{\beta-1}{\alpha}, \forall r\in\mathcal{R}_2.$$
		%    Consequently, according to Lemma \ref{Lem:3}, we can obtain that $\max_{r\in\mathcal{R}_2}g_2(r)=g_2(\lambda_{2})$.
		Under this circumstance, we can obtain that
		$\beta<\sqrt{(\alpha \lambda_{i})^2+2\alpha\lambda_{i}}, \forall i\in\{2,\dots,n\}$ which leads to
		$$\lambda_{i}>\frac{\sqrt{1+\beta^2}-1}{\alpha}>\frac{\beta-1}{\alpha}, \forall i\in\{2,\dots,n\}$$
		and $\sup_{\upsilon\in\mathbb{R}}\delta_{i}(\upsilon)=g_2{(\lambda_{i})}, \forall i\in\{2,\dots,n\}$.
		Due to $\frac{\beta-1}{\alpha}<\lambda_2\leq\cdots\leq\lambda_n$ and $g_2(t)$ is decreasing on $(\frac{\beta-1}{\alpha},+\infty)$, we have $\max_{i=2,\dots,n}g_2{(\lambda_{i})}=g_2(\lambda_2)$.
		Thus, \eqref{eq:A-1} becomes
		\begin{equation*}
			\begin{split}
				\|T_1(s)\|_\infty&=\max_{i=2,\dots,n}\sqrt{g_2{(\lambda_{i})}}\\
				&=\sqrt{g_2(\lambda_2)} \\
				&=\frac{1}{\sqrt{(\alpha\lambda_2)^2-\Big[\sqrt{(\alpha\lambda_2+1)^2-\beta^2}-1\Big]^2}}.
			\end{split}
		\end{equation*}
		
		(\uppercase\expandafter{\romannumeral3})
		$\mathcal{R}_1\neq\emptyset$ and $\mathcal{R}_2\neq\emptyset$;
		
		In this case, there must exists an eigenvalue $\lambda_l~(2\leq l\leq n-1)$ of $L$ such that $\mathcal{R}_1=\{\lambda_2,\dots,\lambda_l\}$ and $\mathcal{R}_2=\{\lambda_{l+1},\dots,\lambda_n\}$.
		Thus, we can get
		$\beta\geq\sqrt{(\alpha \lambda_{i})^2+2\alpha\lambda_{i}}, \forall i\in\{2,\dots,l\}$ and $\beta<\sqrt{(\alpha \lambda_{i})^2+2\alpha\lambda_{i}}, \forall i\in\{l+1,\dots,n\}$ which respectively imply that $\sup_{\upsilon\in\mathbb{R}}\delta_{i}(\upsilon)=g_1{(\lambda_{i})}, \forall i\in\{2,\dots,l\}$ and $\sup_{\upsilon\in\mathbb{R}}\delta_{i}(\upsilon)=g_2{(\lambda_{i})}, \forall i\in\{l+1,\dots,n\}$.
		%    Thus, we can obtain that
		%    \begin{equation}%\label{eq:case1}
			%    	\sup_{\upsilon\in\mathbb{R}}\delta_{i}(\upsilon)=g_1{(\lambda_{i})}, \forall i\in\{2,\dots,l\},
			%    \end{equation}
		%    and
		%    \begin{equation}%\label{eq:case1}
			%    	\sup_{\upsilon\in\mathbb{R}}\delta_{i}(\upsilon)=g_2{(\lambda_{i})}, \forall i\in\{l+1,\dots,n\}.
			%    \end{equation}
		Moreover, it is inferred from $\beta<\sqrt{(\alpha \lambda_{i})^2+2\alpha\lambda_{i}}, \forall i\in\{l+1,\dots,n\}$ that
		$$\lambda_{i}>\frac{\sqrt{1+\beta^2}-1}{\alpha}>\frac{\beta-1}{\alpha}, \forall i\in\{l+1,\dots,n\}.$$
		Since $0<\lambda_2\leq\cdots\leq\lambda_l$ and  $\frac{\beta-1}{\alpha}<\lambda_{l+1}\leq\cdots\leq\lambda_n$,
		it follows from Lemma \ref{Lem:3} that $\max_{i=2,\dots,l}g_1{(\lambda_{i})}=g_1(\lambda_2)$ and $\max_{i=l+1,\dots,n}g_2{(\lambda_{i})}=g_2(\lambda_{l+1})$.
		%    According to Lemma \ref{Lem:3} and \eqref{eq:R2}, one can obtain that $\max_{r\in\mathcal{R}_1}g_1(r)=g_1(\lambda_{2})$ and $\max_{r\in\mathcal{R}_2}g_2(r)=g_2(\lambda_{l+1})$.
		Therefore, based on above facts, \eqref{eq:A-1} can be written as
		\begin{equation*}
			\begin{aligned}
				\|T_1(s)\|_\infty=&\max\bigg\{\max_{i=2,\dots,l}\sqrt{g_1{(\lambda_{i})}},\max_{i=l+1,\dots,n}\sqrt{g_2{(\lambda_{i})}}\bigg\} \\
				=&\max\Big\{\sqrt{g_1(\lambda_{2})},\sqrt{g_2(\lambda_{l+1})}\Big\}
			\end{aligned}
		\end{equation*} 
		Since $\beta\geq\sqrt{(\alpha\lambda_{2})^2+2\alpha\lambda_{2}}$ and $\lambda_{l+1}\geq\lambda_{2}$, it can be readily concluded that
		%    \begin{equation*}
			%    	\max_{r\in\mathcal{R}_1}g_1(r)=\frac{1}{(\alpha\lambda_2)^2}
			%    \end{equation*}
		%    and
		%    \begin{equation*}
			%    	\max_{r\in\mathcal{R}_2}g_2(r)=\frac{1}{(\alpha\lambda_{l+1})^2-\big[\sqrt{(\alpha\lambda_{l+1}+1)^2-\beta^2}-1\big]^2}.
			%    \end{equation*}
		%    It follows from $\beta\geq\sqrt{(\alpha\lambda_{2}+1)^2-1}$ and $\lambda_{l+1}\geq\lambda_{2}$ that
		\begin{equation*}
			\begin{split}
				&g_2(\lambda_{l+1})=\frac{1}{(\alpha\lambda_{l+1})^2-\Big[\sqrt{(\alpha\lambda_{l+1}+1)^2-\beta^2}-1\Big]^2} \\
				&\leq\frac{1}{(\alpha\lambda_{l+1})^2-\Big[\sqrt{(\alpha\lambda_{l+1}+1)^2-(\alpha\lambda_{2}+1)^2+1}-1\Big]^2} \\
				&=\frac{1}{(\alpha\lambda_{2}+1)^2-2+2\sqrt{(\alpha\lambda_{l+1}+1)^2-(\alpha\lambda_{2}+1)^2+1}} \\
				&\leq\frac{1}{(\alpha\lambda_{2}+1)^2} \\
				&<\frac{1}{(\alpha\lambda_2)^2}=g_1(\lambda_{2}).
			\end{split}
		\end{equation*}
		Therefore, we can obtain $\|T_1(s)\|_\infty=\sqrt{g_1(\lambda_{2})}=\frac{1}{\alpha\lambda_2}$.
		
		To sum up above cases, we have $\|T_1(s)\|_\infty=\frac{1}{\alpha\lambda_2}$ as long as $\lambda_2\in\mathcal{R}_1$, i.e., $\beta\geq\sqrt{(\alpha\lambda_{2})^2+2\alpha\lambda_{2}}$. Otherwise, $$\|T_1(s)\|_\infty=\frac{1}{\sqrt{(\alpha\lambda_2)^2-\Big[\sqrt{(\alpha\lambda_2+1)^2-\beta^2}-1\Big]^2}}.$$
		Namely, \eqref{eq:Th1} is derived.\hfill $\square$
	\end{Proof}
	
\begin{figure}[!ht]\centering
		\includegraphics[width=0.48\textwidth,height=0.4\textwidth]{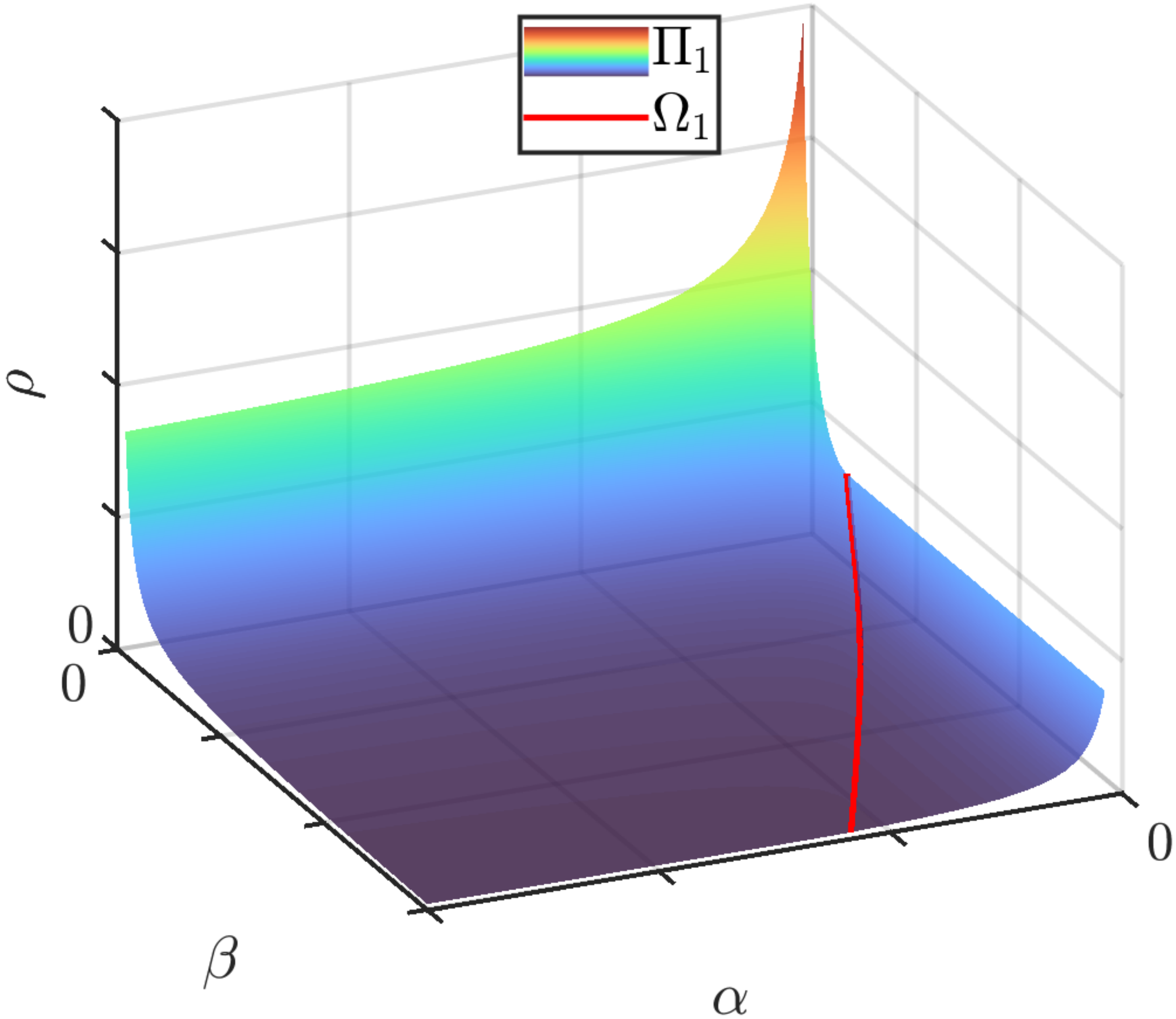}
		\caption{The trends of $\|T_1(s)\|_\infty$ with respect to $\alpha$ and $\beta$.\label{fig2}}
	\end{figure}

	For a connected undirected graph $\mathcal{G}$, $\|T_1(s)\|_\infty$ can be considered as a continuous function of the two variables $\alpha$ and $\beta$. Fig. \ref{fig2} roughly depicts the two-dimensional surface $\Pi_1:=\Big\{(\alpha,\beta,\rho)\in\mathbb{R}^3~\big|~\alpha\in\mathbb{R}^+,\beta\in\mathbb{R}^+,\rho=\|T_1(s)\|_\infty\Big\}$ of $\|T_1(s)\|_\infty$ with respect to $\alpha$ and $\beta$. We use the red solid curve $\Omega_1:=\Big\{(\alpha,\beta,\rho)\in\Pi_1~\big|~\beta=\sqrt{(\alpha\lambda_{2})^2+2\alpha\lambda_{2}}\Big\}$ on $\Pi_1$ to stress the critical condition in \eqref{eq:Th1}.
	%    The red region denotes the part of $\Pi_1$ which is subject to $\beta\geq\sqrt{(\alpha\lambda_{2}+1)^2-1}$, and the blue area is the remainder of $\Pi_1$ constrained by  $\beta<\sqrt{(\alpha\lambda_{2}+1)^2-1}$.
	\subsection{Anti-disturbance capability of the second-order MAS using relative velocity information}
	The following theorem gives the analytic expression of the anti-capability of second-order MAS \eqref{eq:1.1} with relative velocity protocol \eqref{eq:1.3}.
	%Theorem \ref{theorem:2} builds the analytic expression of anti-disturbance capability of the MAS \eqref{eq:1.1} with the relative velocity protocol \eqref{eq:1.3}. 
	%In the subsection, we intend to establish the closed-form expression of $\|T_2(s)\|_\infty$ for the system \eqref{eq:1.8}, which measures the anti-disturbance capability of second-order MAS \eqref{eq:1.1} with the relative velocity protocol \eqref{eq:1.3}. 
	%Provided that the protocol $\mathcal{P}_2$ using relative velocity information is applied to regulate the MAS \eqref{eq:1.1}, the relation among the \Hinf performance, the smallest non-zero eigenvalue of Laplacian matrix and the tunable gain is shown in the following.
	\begin{theorem}\label{theorem:2}
		Consider MAS \eqref{eq:1.1}-\eqref{eq:1.3} in which $G$ is a connected undirected graph with $L$ being its Laplacian matrix.
		Then, we obtain
		%	Consider MAS \eqref{eq:1.1}-\eqref{eq:1.3} in which $\mathcal{G}$ is a connected undirected graph with Laplacian matrix $L$. Then, we obtain
		%	Cosider the MAS \eqref{eq:1.1}-\eqref{eq:1.3} on a connected undirected graph $\mathcal{G}$ with the Laplacian matrix $L$. We can obtain that
		%	Cosider a connected undirected graph $\mathcal{G}$ with the Laplacian matrix $L$. The anti-disturbance capability of MAS \eqref{eq:1.1}-\eqref{eq:1.3} is measured as
		%	Consider a connected undirected graph $\mathcal{G}$ with a Laplacian matrix $L$. The \Hinf performance of the second-order continuous-time MAS \eqref{eq:1.8} is given by
		\begin{equation}\label{eq:Th2}
			\|T_2(s)\|_\infty=\begin{dcases}
				\frac{1}{\sqrt{(\alpha\lambda_2)^2-\Big[\sqrt{(\alpha\lambda_2+1)^2-(\beta\lambda_2)^2}-1\Big]^2}},& \text{if}~~0<\beta<\sqrt{\alpha^2+\frac{2\alpha}{\lambda_2}}, \\
				\frac{1}{\alpha\lambda_2}, & \text{if}~~\beta\geq\sqrt{\alpha^2+\frac{2\alpha}{\lambda_2}}.
			\end{dcases}
		\end{equation}
		%    and assume the form
		%    $0=Re(\lambda_1)< Re(\lambda_2)\leq \cdots \leq Re(\lambda_n).$
	\end{theorem}
	\begin{Proof}
		Since the undirected graph $\mathcal{G}$ is connected, it follows from Lemma \ref{Lem:2} that there exists an orthogonal matrix $Q\in\mathbb{R}^{n\times n}$ such that \eqref{eq:0.2} and \eqref{eq:0.3} hold.
		Then performing the orthogonal transformation \eqref{eq:2.2} for the system \eqref{eq:1.8} yields
		\begin{equation}\label{eq:2.9}
			\begin{dcases}
				\begin{bmatrix}
					\dot{\hat{x}}(t) \\ \dot{\hat{v}}(t)
				\end{bmatrix}=\begin{bmatrix}
					\mathbf{0} & I_{n} \\ -\alpha \bar{L} & -\beta \bar{L}
				\end{bmatrix}\begin{bmatrix}
					\hat{x}(t) \\ \hat{v}(t)
				\end{bmatrix}+\begin{bmatrix}
					\mathbf{0} \\ I_{n}
				\end{bmatrix}\hat{\omega}(t), \\
				\hat{y}(t)=\begin{bmatrix}
					\bar{\Phi}_n & \mathbf{0} \\ \mathbf{0} & \bar{\Phi}_n
				\end{bmatrix}\begin{bmatrix}
					\hat{x}(t) \\ \hat{v}(t)
				\end{bmatrix},
			\end{dcases}
		\end{equation}
		where $$\bar{L}=\begin{bmatrix}
			\bar{L}_1 & \mathbf{0}_{n-1} \\
			\mathbf{0}_{n-1}^{\top} & 0
		\end{bmatrix}.$$
		It is observed from Lemma \ref{Lem:1} and \eqref{eq:0.3} that positive definite matrix $\bar{L}_1$ owns the same non-zero eigenvalues of $L$ which implies that $-\bar{L}_1$ is Hurwitz stable.
		Obviously, the system \eqref{eq:2.9} is decomposed into an observable asymptotically stable subsystem of order $2n-2$ and an unobservable marginally stable subsystem of order $2$.
		According to \cite{Siami2015}, $\|T_2(s)\|_\infty$ is well-defined and completely reliant on the asymptotically stable subsystem. Thus we turn to analyze the asymptotically stable subsystem of \eqref{eq:2.9} which is given as follows
		\begin{equation}\label{eq:2.9-1}
			\begin{dcases}
				\begin{bmatrix}
					\dot{\hat{x}}^1(t) \\ \dot{\hat{v}}^1(t)
				\end{bmatrix}=\begin{bmatrix}
					\mathbf{0} & I_{n-1}  \\ -\alpha\bar{L}_1 & -\beta\bar{L}_1
				\end{bmatrix}
				\begin{bmatrix}
					\hat{x}^1(t) \\ \hat{v}^1(t)
				\end{bmatrix}+\begin{bmatrix}
					\mathbf{0} \\ I_{n-1}
				\end{bmatrix}\hat{\omega}^1(t), \\
				\hat{y}^1(t)=\begin{bmatrix}
					I_{n-1} & \mathbf{0}\\ \mathbf{0} &  I_{n-1}
				\end{bmatrix}\begin{bmatrix}
					\hat{x}^1(t) \\ \hat{v}^1(t)
				\end{bmatrix},
			\end{dcases}
		\end{equation}
		where $\hat{x}^1(t)=\begin{bmatrix}I_{n-1} & \mathbf{0}_{n-1}\end{bmatrix}\hat{x}(t)$, $\hat{v}^1(t)=\begin{bmatrix}I_{n-1} & \mathbf{0}_{n-1}\end{bmatrix}$\ \ \ $\hat{v}(t)$ and $\hat{\omega}^1(t)=\begin{bmatrix}I_{n-1} & \mathbf{0}_{n-1}\end{bmatrix}\hat{\omega}(t).$
		Since $\bar{L}_1$ is positive definite, there must exists an orthogonal matrix $V\in\mathbb{R}^{(n-1)\times (n-1)}$ such that $V^\top\bar{L}_1V=\varLambda$, where $\varLambda=\text{diag}\{\lambda_2,\lambda_3,\dots,\lambda_n\}$.
		Then applying the orthogonal transformation \eqref{eq:2.7} for system \eqref{eq:2.9-1} elicits the following system
		\begin{equation}\label{eq:2.10}
			\begin{dcases}
				\begin{bmatrix}
					\dot{\tilde{x}}(t) \\ \dot{\tilde{v}}(t)
				\end{bmatrix}=\begin{bmatrix}
					\mathbf{0} & I_{n-1}  \\ -\alpha\varLambda & -\beta\varLambda
				\end{bmatrix}
				\begin{bmatrix}
					\tilde{x}(t) \\ \tilde{v}(t)
				\end{bmatrix}+\begin{bmatrix}
					\mathbf{0} \\ I_{n-1}
				\end{bmatrix}\tilde{\omega}(t), \\
				\tilde{y}(t)=\begin{bmatrix}
					I_{n-1} & \mathbf{0}\\ \mathbf{0} &  I_{n-1}
				\end{bmatrix}\begin{bmatrix}
					\tilde{x}(t) \\ \tilde{v}(t)
				\end{bmatrix}.
			\end{dcases}
		\end{equation}
		We indicate the transfer matrix of the system \eqref{eq:2.9}, \eqref{eq:2.9-1} and \eqref{eq:2.10} by $T_6(s)$, $T_7(s)$ and $T_8(s)$, respectively. A simple calculation gives rise to
		\begin{equation}\label{eq:T2=T8}
			\|T_2(s)\|_\infty=\|T_6(s)\|_\infty=\|T_7(s)\|_\infty=\|T_8(s)\|_\infty.
		\end{equation}
		Next, we intend to compute $\|T_8(s)\|_\infty$.
		
		The transfer matrix $T_8(s)$ is given as
		\begin{equation*}
			T_8(s)=\left[\begin{array}{ccc}
				\Psi_2& &\\
				& \ddots &\\
				& & \Psi_n \\
				\hdashline
				\Omega_2& &\\
				& \ddots &\\
				& & \Omega_n
			\end{array}\right],
		\end{equation*}
		where $\Psi_i=\dfrac{1}{s^2+\beta\lambda_i s+\alpha \lambda_i}$ and $\Omega_i=\dfrac{s}{s^2+\beta\lambda_i s+\alpha \lambda_i}$, $i=2,\dots,n$.
		We can obtain that
		\begin{equation*}
			T_8^H(\mathbf{j}\upsilon)T_8(\mathbf{j}\upsilon)=\begin{bmatrix}
				\theta_2(\upsilon)& &\\
				& \ddots &\\
				& & \theta_n(\upsilon)
			\end{bmatrix}
		\end{equation*}
		is diagonal, where $$\theta_i(\upsilon)=\frac{1+\upsilon^2}{(\alpha\lambda_i-\upsilon^2)^2+(\beta\lambda_i \upsilon)^2}>0,~i=2,\dots,n,~\upsilon\in\mathbb{R}.$$
		By inspecting Definition \ref{Def:1} and \eqref{eq:T2=T8}, $\|T_2(s)\|_\infty$ can be written as
		\begin{equation}\label{eq:B-1}
			\begin{split}
				\|T_2(s)\|_\infty&=\|T_8(s)\|_\infty \\
				&=\sup_{\upsilon\in\mathbb{R}} 
				\sqrt{\lambda_{max}\big[T_8^H(\mathbf{j}\upsilon)T_8(\mathbf{j}\upsilon)\big]} \\
				&=\max_{i=2,\dots,n}\sqrt{\sup_{\upsilon\in\mathbb{R}}\theta_{i}(\upsilon)}.
			\end{split}
		\end{equation}
		It follows from 
		\begin{equation}\label{eq:diff_theta_i}
			\frac{\mathrm{d}\theta_i(\upsilon)}{\mathrm{d}\upsilon}=\frac{-2\upsilon\big[\upsilon^4+2\upsilon^2+\beta^2\lambda_{i}^2-(\alpha\lambda_{i})^2-2\alpha\lambda_{i}\big]}{\big[\upsilon^4+(\beta^2\lambda_{i}^2-2\alpha\lambda_{i})\upsilon^2+(\alpha\lambda_{i})^2\big]^2}=0
		\end{equation}
		that
		\begin{equation}\label{eq:numerator_diff_theta}
			\upsilon\big[\upsilon^4+2\upsilon^2+\beta^2\lambda_{i}^2-(\alpha\lambda_{i})^2-2\alpha\lambda_{i}\big]=0.
		\end{equation}  
		Obviously, $\sup_{\upsilon\in\mathbb{R}}\theta_{i}(\upsilon)$ is determined by $\beta^2\lambda_i^2-2\alpha\lambda_i-\alpha^2\lambda_i^2$.
		If $\beta\geq\sqrt{\alpha^2+\frac{2\alpha}{\lambda_i}}$, by solving \eqref{eq:numerator_diff_theta}, we get
		\begin{equation*}
			\sup_{\upsilon\in\mathbb{R}}\theta_{i}(\upsilon)=\theta_{i}(\upsilon^*_{i,1})=g_1{(\lambda_{i})}=\frac{1}{(\alpha\lambda_i)^2},
		\end{equation*}
		where $\upsilon^*_{i,1}=0$.
		If $\beta<\sqrt{\alpha^2+\frac{2\alpha}{\lambda_i}}$, by solving \eqref{eq:numerator_diff_theta}, we obtain
		\begin{equation*}
			\begin{aligned}
				\sup_{\upsilon\in\mathbb{R}}\theta_{i}(\upsilon)&=\theta_{i}(\upsilon^*_{i,2})=\theta_{i}(\upsilon^*_{i,3})=g_3{(\lambda_{i})}\\
				&=\frac{1}{(\alpha\lambda_i)^2-(\sqrt{\Theta_i}-1)^2},
			\end{aligned}
		\end{equation*}
		where $\upsilon^*_{i,2},\upsilon^*_{i,3}=\pm\sqrt{\sqrt{\Theta_i}-1}$, and $\Theta_i=(\alpha\lambda_i+1)^2-(\beta\lambda_i)^2$.
		The trends of the function $\theta_{i}(\upsilon)$ under the conditions $\beta\geq\sqrt{\alpha^2+\frac{2\alpha}{\lambda_i}}$ and $\beta<\sqrt{\alpha^2+\frac{2\alpha}{\lambda_i}}$ resemble the Fig. \ref{fig1}. For saving space, we omit it here.
		%    Nevertheless, as a result of
		%    $$\frac{\mathrm{d}\theta_i(\upsilon)}{\mathrm{d}\upsilon}=0 \Rightarrow \upsilon\big[\upsilon^4+2\upsilon^2+(\beta^2\lambda_i^2-2\alpha\lambda_i-\alpha^2\lambda_i^2)\big]=0,$$
		%    the function $\theta_i(\upsilon)$ exhibits different trends in terms of $\beta^2\lambda_i^2-2\alpha\lambda_i-\alpha^2\lambda_i^2$. Then $\theta_i(\upsilon)$ can be denoted as
		%    \begin{equation*}
			%	\begin{dcases}
				%		\theta_i(\upsilon)=\theta_{i,1}(\upsilon)>0 & \text{if}~\beta\geq\sqrt{\alpha^2+\frac{2\alpha}{\lambda_i}}, \\
				%		\theta_i(\upsilon)=\theta_{i,2}(\upsilon)>0 & \text{if}~\beta<\sqrt{\alpha^2+\frac{2\alpha}{\lambda_i}},
				%	\end{dcases}
			%    \end{equation*}
		%whose trends resemble the Fig. \ref{fig1}. By solving $\dfrac{\mathrm{d}\theta_i(\upsilon)}{\mathrm{d}\upsilon}=0$ under above two cases, we obtain
		%\begin{equation*}
		%	\sup_{\upsilon\in\mathbb{R}}\theta_{i,1}(\upsilon)=\theta_{i,1}(\upsilon^*_{i,1})=\frac{1}{(\alpha\lambda_i)^2}, 
		%\end{equation*}
		%and
		%\begin{equation*}
		%	\begin{aligned}
			%		\sup_{\upsilon\in\mathbb{R}}\theta_{i,2}(\upsilon)&=\theta_{i,2}(\upsilon^*_{i,2})=\theta_{i,2}(\upsilon^*_{i,3})=\frac{1}{(\alpha\lambda_i)^2-(\sqrt{\Theta_i}-1)^2} \\
			%        &>\theta_{i,2}(\upsilon^*_{i,1})=\frac{1}{(\alpha\lambda_i)^2},
			%	\end{aligned}
		%\end{equation*}
		%where $\upsilon^*_{i,1}=0$, $\upsilon^*_{i,2},\upsilon^*_{i,3}=\pm\sqrt{-1+\sqrt{\Theta_i}}$ and $\Theta_i=(\alpha\lambda_i+1)^2-(\beta\lambda_i)^2$.
		
		Aforementioned analysis prompts us to dictate two sets of non-zero eigenvalues of $L$, which are given as
		\begin{equation*}
			\mathcal{K}_1=\Bigg\{k\in\Gamma~\bigg|~\beta\geq\sqrt{\alpha^2+\frac{2\alpha}{k}}\Bigg\}
		\end{equation*}
		and
		\begin{equation*}
			\mathcal{K}_2=\Bigg\{k\in\Gamma~\bigg|~\beta<\sqrt{\alpha^2+\frac{2\alpha}{k}}\Bigg\}.
		\end{equation*}
		Obviously, we can get $\mathcal{K}_1\cup\mathcal{K}_2=\Gamma$ and  
		$\mathcal{K}_1\cap\mathcal{K}_2=\emptyset$.
		We will complete the proof by enumeration.
		
		(\uppercase\expandafter{\romannumeral1})
		$\mathcal{K}_1=\Gamma$ and $\mathcal{K}_2=\emptyset$;
		
		In this case, we have $\beta\geq\sqrt{\alpha^2+\frac{2\alpha}{\lambda_{i}}}, \forall i\in\{2,\dots,n\}$ which implies that $\sup_{\upsilon\in\mathbb{R}}\theta_{i}(\upsilon)=g_1{(\lambda_{i})}, \forall i\in\{2,\dots,n\}$. 
		Because $0<\lambda_2\leq\cdots\leq\lambda_n$ and $g_1{(t)}$ is decreasing on $(0,+\infty)$, one can obtain that $\max_{i=2,\dots,n}g_1{(\lambda_{i})}=g_1(\lambda_2)$.
		Then \eqref{eq:B-1} becomes
		\begin{equation*}
			\|T_2(s)\|_\infty=\max_{i=2,\dots,n}\sqrt{g_1{(\lambda_{i})}}=\sqrt{g_1(\lambda_2)}=\frac{1}{\alpha\lambda_2}.
		\end{equation*}
		
		(\uppercase\expandafter{\romannumeral2})
		$\mathcal{K}_1=\emptyset$ and $\mathcal{K}_2=\Gamma$;
		
		%It should be stressed that if 
		%$\beta>\alpha$, then we have
		%\begin{equation}\label{eq:beta>alpha}
		%	0<k<\dfrac{2\alpha}{\beta^2-\alpha^2}, \forall k\in\mathcal{K}_2.
		%\end{equation}
		
		In this case, we get $\beta<\sqrt{\alpha^2+\frac{2\alpha}{\lambda_{i}}}, \forall i\in\{2,\dots,n\}$ which leads to $\sup_{\upsilon\in\mathbb{R}}\theta_{i}(\upsilon)=g_3{(\lambda_{i})}, \forall i\in\{2,\dots,n\}$.
		If $\beta\leq\alpha$, it is inferred from $0<\lambda_2\leq\cdots\leq\lambda_n$ and Lemma \ref{Lem:3} that
		\begin{equation}\label{eq:max_g2}
			\max_{i=2,\dots,n}g_3{(\lambda_{i})}=g_3(\lambda_2).
		\end{equation}
		If $\beta>\alpha$, it follows from $0<\lambda_2\leq\cdots\leq\lambda_n$ and $\beta<\sqrt{\alpha^2+\frac{2\alpha}{\lambda_{i}}}, \forall i\in\{2,\dots,n\}$ that $0<\lambda_2\leq\cdots\leq\lambda_n<\frac{2\alpha}{\beta^2-\alpha^2}$.
		Then according to Lemma \ref{Lem:3}, equation \eqref{eq:max_g2} still holds.
		%To sum up, \eqref{eq:max_g2} holds for any $\alpha$ and $\beta$ in case (\uppercase\expandafter{\romannumeral2}).
		%That is to say, \eqref{eq:max_g2} always holds as long as $\mathcal{K}_2=\Gamma$.
		Building on above analysis, no matter $\beta\leq\alpha$ or $\beta>\alpha$, \eqref{eq:B-1} can be written as
		\begin{equation*}
			\begin{split}
				\|T_2(s)\|_\infty&=\max_{i=2,\dots,n}\sqrt{g_3{(\lambda_{i})}} \\
				&=\sqrt{g_3(\lambda_2)} \\
				&=\frac{1}{\sqrt{(\alpha\lambda_2)^2-\Big[\sqrt{(\alpha\lambda_2+1)^2-(\beta\lambda_{2})^2}-1\Big]^2}}.
			\end{split}
		\end{equation*}
		
		(\uppercase\expandafter{\romannumeral3})
		$\mathcal{K}_1\neq\emptyset$ and $\mathcal{K}_2\neq\emptyset$.
		
		Under this circumstance, there must exists an eigenvalue $\lambda_{m}$ $(2\leq m\leq n-1)$ of $L$ such that $\mathcal{K}_1=\{\lambda_{m+1},\dots,\lambda_n\}$ and $\mathcal{K}_2=\{\lambda_{2},\dots,\lambda_m\}$.
		Then we have $\beta\geq\sqrt{\alpha^2+\frac{2\alpha}{\lambda_{i}}}, \forall i\in\{m+1,\dots,n\}$ and $\beta<\sqrt{\alpha^2+\frac{2\alpha}{\lambda_{i}}}, \forall i\in\{2,\dots,m\}$ which respectively lead to
		$\sup_{\upsilon\in\mathbb{R}}\theta_{i}(\upsilon)=g_1{(\lambda_{i})}, \forall i\in\{m+1,\dots,n\}$ and $\sup_{\upsilon\in\mathbb{R}}\theta_{i}(\upsilon)=g_3{(\lambda_{i})}, \forall i\in\{2,\dots,m\}$.
		Since $0<\lambda_{m+1}\leq\cdots\leq\lambda_n$ and $g_1{(t)}$ is decreasing on $(0,+\infty)$, one can obtain that $\max_{i=m+1,\dots,n}g_1{(\lambda_{i})}=g_1(\lambda_{m+1})$.
		Furthermore, it can be readily concluded that
		\begin{equation*}
			\alpha<\sqrt{\alpha^2+\dfrac{2\alpha}{\lambda_{n}}}\leq\dots\leq\sqrt{\alpha^2+\dfrac{2\alpha}{\lambda_{m+1}}}\leq\beta.
		\end{equation*}
		Since $\alpha<\beta$, it follows from $0<\lambda_2\leq\cdots\leq\lambda_m$ and $\beta<\sqrt{\alpha^2+\frac{2\alpha}{\lambda_{i}}}, \forall i\in\{2,\dots,m\}$ that $0<\lambda_2\leq\cdots\leq\lambda_m<\frac{2\alpha}{\beta^2-\alpha^2}$.
		According to Lemma \ref{Lem:3}, we can obtain  $\max_{i=2,\dots,m}g_3{(\lambda_{i})}=g_3(\lambda_2)$.
		
		Based on above analysis, \eqref{eq:B-1} can be written as
		\begin{equation*}
			\begin{aligned}
				\|T_2(s)\|_\infty=&\max\bigg\{\max_{i=m+1,\dots,n}\sqrt{g_1{(\lambda_{i})}},\max_{i=2,\dots,m}\sqrt{g_3{(\lambda_{i})}}\bigg\} \\
				=&\max\Big\{\sqrt{g_1(\lambda_{m+1})},\sqrt{g_3(\lambda_{2})}\Big\}
			\end{aligned}
		\end{equation*} 
		It follows from $\beta<\sqrt{\alpha^2+\frac{2\alpha}{\lambda_{2}}}$ and $\lambda_{m+1}\geq\lambda_{2}$ that
		\begin{equation*}
			\begin{split}
				g_3(\lambda_{2})=&\frac{1}{(\alpha\lambda_{2})^2-\Big[\sqrt{(\alpha\lambda_{2}+1)^2-(\beta\lambda_{2})^2}-1\Big]^2} \\
				>&\frac{1}{(\alpha\lambda_{2})^2}\geq\frac{1}{(\alpha\lambda_{m+1})^2}=g_1(\lambda_{m+1}).
			\end{split}
		\end{equation*}
		Therefore, we can obtain 
		\begin{equation*}
			\|T_2(s)\|_\infty=\sqrt{g_3(\lambda_{2})}
			=\frac{1}{\sqrt{(\alpha\lambda_{2})^2-\Big[\sqrt{(\alpha\lambda_{2}+1)^2-(\beta\lambda_{2})^2}-1\Big]^2}}
		\end{equation*}

		To summarize above cases, we can conclude that $$\|T_2(s)\|_\infty=\frac{1}{\sqrt{(\alpha\lambda_{2})^2-\Big[\sqrt{(\alpha\lambda_{2}+1)^2-(\beta\lambda_{2})^2}-1\Big]^2}}$$
		as long as $\lambda_2\in\mathcal{K}_2$, i.e., $\beta<\sqrt{\alpha^2+\frac{2\alpha}{\lambda_{2}}}$. Otherwise, $\|T_2(s)\|_\infty=\frac{1}{\alpha\lambda_2}$. That is to say, \eqref{eq:Th2} is obtained.\hfill $\square$
	\end{Proof}
	
\begin{figure}[!ht]\centering
		\includegraphics[width=0.48\textwidth,height=0.4\textwidth]{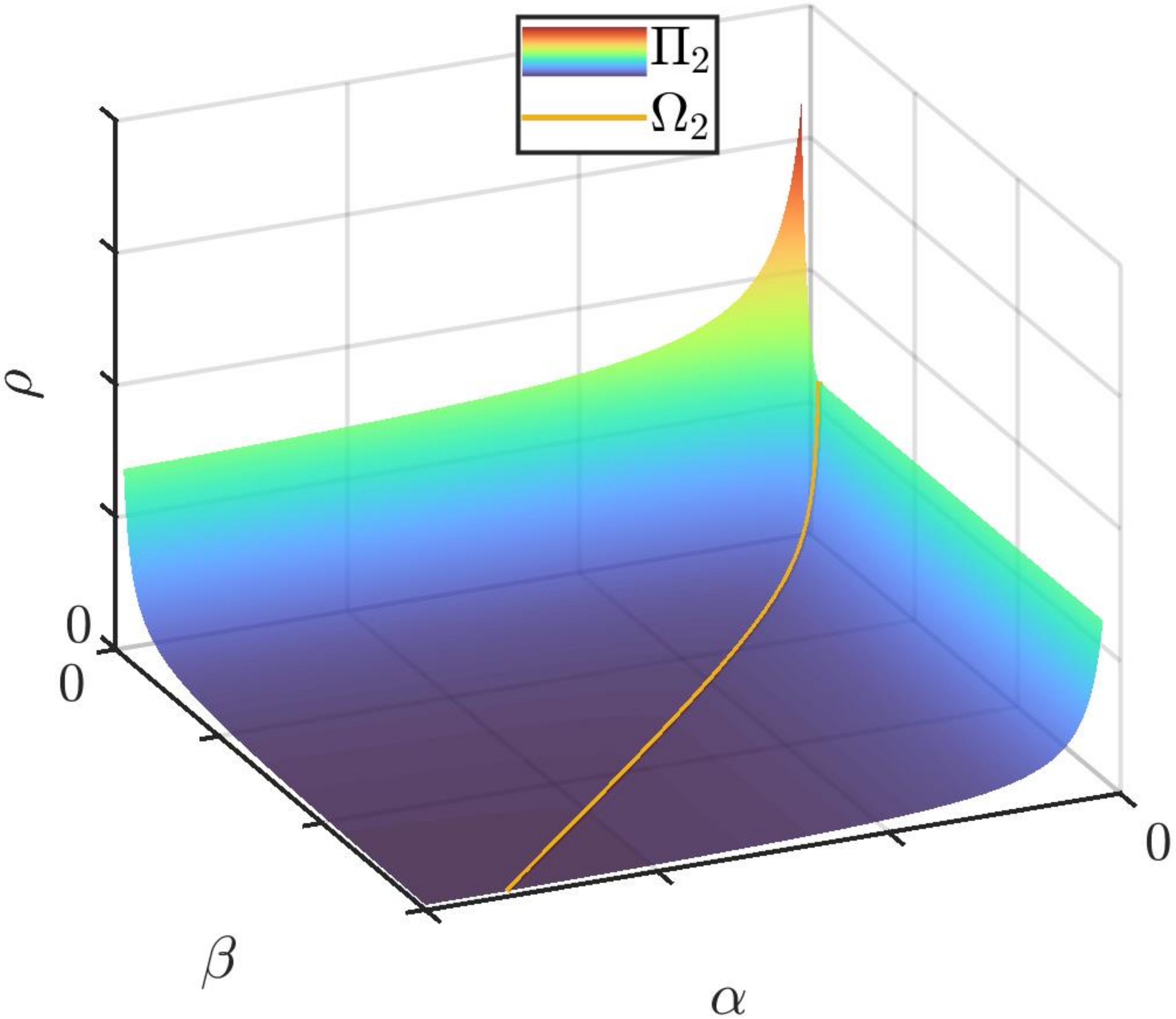}
		\caption{The trends of $\|T_2(s)\|_\infty$ with respect to $\alpha$ and $\beta$.\label{fig3}}
	\end{figure}

	In fact, $\|T_2(s)\|_\infty$ can be deemed as a continuous function of the two variables $\alpha\in\mathbb{R}^+$ and $\beta\in\mathbb{R}^+$ on a given connected undirected graph $\mathcal{G}$.
	We roughly portray the two-dimensional surface $\Pi_2:=\Big\{(\alpha,\beta,\rho)\in\mathbb{R}^3~\big|~\alpha\in\mathbb{R}^+,\beta\in\mathbb{R}^+,\rho=\|T_2(s)\|_\infty\Big\}$ of $\|T_2(s)\|_\infty$ with respect to $\alpha$ and $\beta$ in Fig. \ref{fig3}. As one can see from Fig. \ref{fig3}, $\Pi_2$ is seperated into two regions by the yellow solid curve $$\Omega_2:=\Bigg\{(\alpha,\beta,\rho)\in\Pi_2~\bigg|~\beta=\sqrt{\alpha^2+\dfrac{2\alpha}{\lambda_2}}\Bigg\},$$
	which refers to the critical condition in \eqref{eq:Th2}.
	%Particularly, the red region indicates the part of $\Pi_1$ which is subject to $\beta\geq\sqrt{\alpha^2+\dfrac{2\alpha}{\lambda_2}}$, and the blue area represents the other part of $\Pi_2$ restricted to $\beta<\sqrt{\alpha^2+\dfrac{2\alpha}{\lambda_2}}$.
	\begin{remark}
		It is observed from Theorem \ref{theorem:1} and Theorem \ref{theorem:2} that $\|T_1(s)\|_\infty$ and $\|T_2(s)\|_\infty$ are both inversely proportional to $\lambda_{2}$ for any given tunable gains, which is consistent with the result for first-order MASs \cite{Siami2014}. In other words, optimizing networks to generate a larger $\lambda_{2}$ is still valid to improve the anti-disturbance capability for second-order MASs.
	\end{remark}
	\subsection{Protocol selection for better anti-disturbance capability}
	According to the analytic expressions presented in Theorems \ref{theorem:1} and \ref{theorem:2}, in Theorem \ref{theorem:3} we will establish the graph conditions of protocol selection for better anti-disturbance capability.
	%We propose a graph condition in the following theorem for the protocol selection between the absolute velocity protocol \eqref{eq:1.2} and the relative velocity protocol \eqref{eq:1.3} to obtain better anti-disturbance capability for MAS \eqref{eq:1.1}.
	\begin{theorem}\label{theorem:3}
		Consider the second-order MAS \eqref{eq:1.1} on a connected undirected graph $\mathcal{G}$ with Laplacian matrix $L$. Then, we conclude that
		\begin{itemize}
			\item[1)] The protocol \eqref{eq:1.2} outperforms the protocol \eqref{eq:1.3} if $\lambda_{2}<1$$;$
			\item[2)] The protocol \eqref{eq:1.3} outperforms the protocol \eqref{eq:1.2} if $\lambda_{2}>1$$;$
			\item[3)] The protocol \eqref{eq:1.2} performs as well as the protocol \eqref{eq:1.3} if $\lambda_{2}=1$$.$
		\end{itemize}
		%    \begin{equation*}
			%    	\begin{cases}
				%    		\eqref{eq:1.2} \succ \eqref{eq:1.3}, & \text{if}~\lambda_{2}<1, \\
				%    		\eqref{eq:1.2} \sim \eqref{eq:1.3}, & \text{if}~\lambda_{2}=1, \\
				%    		\eqref{eq:1.3} \succ \eqref{eq:1.2}, & \text{if}~\lambda_{2}>1,
				%    	\end{cases}
			%    \end{equation*}
	\end{theorem}
	\begin{Proof}
		Firstly, we intend to prove the conclusion 3). By substituting $\lambda_{2}=1$ into \eqref{eq:Th1} and \eqref{eq:Th2}, respectively, it is easy to verify that $\|T_1(s)\|_\infty\equiv\|T_2(s)\|_\infty$ for any $\alpha>0$ and $\beta>0$. Therefore, according to Definition \ref{Def:3}, we can say that the protocol \eqref{eq:1.2} performs as well as the protocol \eqref{eq:1.3}.
		
		Next, we will prove the conclusion 1).
		Let $p=\sqrt{(\alpha\lambda_2)^2+2\alpha\lambda_{2}}$ and $q=\sqrt{\alpha^2+\frac{2\alpha}{\lambda_2}}$. Note that $p=\lambda_{2}q$.
		Recall that $\Delta_2=(\alpha\lambda_2+1)^2-\beta^2$ and $\Theta_2=(\alpha\lambda_2+1)^2-(\beta\lambda_2)^2$.
		It follows from $\lambda_{2}<1$ that $p<q$ and $\Delta_2<\Theta_2$.
		We will complete the proof of conclusion 1) by enumeration.
		
		(\uppercase\expandafter{\romannumeral1})
		$\beta<p<q$;
		
		As seen in Theorem \ref{theorem:1} and Theorem \ref{theorem:2}, we can obtain
		$$\|T_1(s)\|_\infty=\frac{1}{\sqrt{(\alpha\lambda_2)^2-(\sqrt{\Delta_2}-1)^2}}$$
		and 
		$$\|T_2(s)\|_\infty=\frac{1}{\sqrt{(\alpha\lambda_2)^2-(\sqrt{\Theta_2}-1)^2}}.$$
		Since $\beta<p$ and $\Delta_2<\Theta_2$, we have $1<\Delta_2<\Theta_2$.
		It follows that
		\begin{equation*}
			\sqrt{(\alpha\lambda_2)^2-(\sqrt{\Theta_2}-1)^2}<\sqrt{(\alpha\lambda_2)^2-(\sqrt{\Delta_2}-1)^2}
		\end{equation*}
		which means that $\|T_1(s)\|_\infty<\|T_2(s)\|_\infty$.
		
		(\uppercase\expandafter{\romannumeral2})
		$p\leq\beta<q$;
		
		In terms of Theorem \ref{theorem:1} and Theorem \ref{theorem:2}, one can obtain that $\|T_1(s)\|_\infty=\frac{1}{\alpha\lambda_2}$ and $$\|T_2(s)\|_\infty=\frac{1}{\sqrt{(\alpha\lambda_2)^2-(\sqrt{\Theta_2}-1)^2}}.$$
		It is inferred from
		\begin{equation*}
			\sqrt{(\alpha\lambda_2)^2-(\sqrt{\Theta_2}-1)^2}<\sqrt{(\alpha\lambda_2)^2}
		\end{equation*}
		that $\|T_1(s)\|_\infty<\|T_2(s)\|_\infty$.
		
		(\uppercase\expandafter{\romannumeral3})
		$p<q\leq\beta$.
		
		It follows from Theorem \ref{theorem:1} and Theorem \ref{theorem:2} that $\|T_1(s)\|_\infty=\|T_2(s)\|_\infty=\frac{1}{\alpha\lambda_2}$.
		
		In summary, $\|T_1(s)\|_\infty\leq\|T_2(s)\|_\infty$ holds for any $\alpha>0$ and $\beta>0$ when $\lambda_{2}<1$. 
		Therefore, as stated in Definition \ref{Def:3}, we can say that the protocol \eqref{eq:1.2} outperforms the protocol \eqref{eq:1.3} if $\lambda_{2}<1$.
		Similarly, if $\lambda_{2}>1$, we can also conclude that the protocol \eqref{eq:1.3} outperforms the protocol \eqref{eq:1.2}. For saving space, we omit it.\hfill $\square$
	\end{Proof}
	
	%    In the light of Definition \ref{Def:1} and Theorem \ref{theorem:3}, we can conclude the following statement.
	%    \begin{proposition}\label{proposition:1}
		%    For the second-order MAS \eqref{eq:1.1} on the connected graph $\mathcal{G}$,
		%    
		%    $1)$ $\mathcal{P}_1\succ\mathcal{P}_2$ if $\lambda_{2}<1$$;$
		%    
		%    $2)$ $\mathcal{P}_2\succ\mathcal{P}_1$ if $\lambda_{2}>1$$;$
		%    
		%    $3)$ $\mathcal{P}_1\sim\mathcal{P}_2$ if $\lambda_{2}=1$.
		%    \end{proposition}
	As one can see, the protocol selection for better anti-disturbance capability exclusively relies on the graph-theoretic feature $\lambda_{2}$. 
	It means that we do not have to execute complicated calculation and comparison on $\|T_1(s)\|_\infty$ and $\|T_2(s)\|_\infty$ for all positive tunable gains $\alpha$ and $\beta$ as Definition \ref{Def:3}, the graph conditions about $\lambda_{2}$ are more consice and tractable.
	
	Additionally, our results embody a twofold approach for improving the anti-disturbance capability of MAS \eqref{eq:1.1}. In a fixed communication network scenario, we can first opt for a better communication protocol in terms of Theorem \ref{theorem:3}. 
	Then the importance of the tunable gains $\alpha$ and $\beta$ now comes to the fore. 
	As shown in \eqref{eq:Th1} and \eqref{eq:Th2}, $\|T_1(s)\|_\infty$ and $\|T_2(s)\|_\infty$ can be viewed as the functions of tunable gains $\alpha$ and $\beta$. 
	Note that the partial derivatives $\frac{\partial \|T_1(s)\|_\infty}{\partial \alpha}$, $\frac{\partial \|T_1(s)\|_\infty}{\partial \beta}$, $\frac{\partial \|T_2(s)\|_\infty}{\partial \alpha}$ and $\frac{\partial \|T_2(s)\|_\infty}{\partial \beta}$ are all continuous, and we have
	\begin{equation*}
		\begin{dcases}
			\dfrac{\partial \|T_1(s)\|_\infty}{\partial \alpha}<0 , \dfrac{\partial \|T_2(s)\|_\infty}{\partial \alpha}<0, \\
			\dfrac{\partial \|T_1(s)\|_\infty}{\partial \beta}\leq0, \dfrac{\partial \|T_2(s)\|_\infty}{\partial \beta}\leq0,
		\end{dcases}
	\end{equation*}
	where the equations
	$\frac{\partial \|T_1(s)\|_\infty}{\partial \beta}=0$ and $\frac{\partial \|T_2(s)\|_\infty}{\partial \beta}=0$
	hold if and only if $\beta\geq\sqrt{(\alpha\lambda_2)^2+2\alpha\lambda_2}$ and $\beta\geq\sqrt{\alpha^2+\frac{2\alpha}{\lambda_2}}$, respectively.
	Therefore, we are able to further improve the anti-disturbance capability by increasing the tunable gain $\alpha$ or $\beta$, which is also illustrated in Fig. \ref{fig2} and Fig. \ref{fig3}. Nevertheless, on account of
	\begin{equation*}
		\begin{gathered}
			\lim\limits_{\alpha\to\infty}\|T_1(s)\|_\infty=\dfrac{1}{\beta}, \lim\limits_{\beta\to\infty}\|T_1(s)\|_\infty=\dfrac{1}{\alpha\lambda_{2}},\\
			\lim\limits_{\alpha\to\infty}\|T_2(s)\|_\infty=\dfrac{1}{\beta\lambda_{2}}, \lim\limits_{\beta\to\infty}\|T_2(s)\|_\infty=\dfrac{1}{\alpha\lambda_{2}},
		\end{gathered}
	\end{equation*}
	adjusting single tunable gain to improve anti-disturbance capability is limited. Consequently, the optimal anti-disturbance capability is obtained when both tunable gains are sufficiently large.
	\begin{remark}
		The protocol selection approach embodies potential advantages in practical applications.
		Specifically, if the network is unable to rearrange or expand which means that modifying the network to improve the anti-disturbance capability is impracticable, we can still optimize the anti-disturbance capability by manipulating every agents to follow the identical optimal communication protocol.
		Moreover, for a certain control task of a MAS, selecting from existing practicable protocols rather than designing a new protocol is more tractable and highly efficient in the distributed scenario.
	\end{remark}
	
	%%%%%%%%%%%%%%%%%%%%%%%%%%%%%%%%%%%%%%%%%%%%%%%%%%%%%%%%%%%%%%%
	\section{Simlulations}\label{Sec:4}
	%%%%%%%%%%%%%%%%%%%%%%%%%%%%%%%%%%%%%%%%%%%%%%%%%%%%%%%%%%%%%%%
	We provide simulations to illustrate the effectiveness of our theoretical results.
	
	Consider the following well-known graphs with $n\geq2$ vertices and $0$-$1$ edge weights, which include undirected complete graphs
	$K_n$, undirected star graphs $S_n$, undirected path graphs $P_n$, and
	undirected $2k$-regular ring lattices $C_{k,n}$. It should be stressed that $C_{k,n}$ $(n\geq2k+1)$ are highly structured networks with nodes placed on a ring, each connecting to its $2k$ nearest neighbors.
	For more details about those networks, one can refer to \cite{Lewis2014,Wu2007}.
	For above networks, Table \ref{tab-1} gives the analytic expressions of the minimum non-zero eigenvalue and the network density.
	%	The diagrams of these networks are omitted here and one can see more details in \cite{Lewis2014,Wu2007}.
	%	For above networks, Table \ref{tab-1} summarizes the analytic expression of the minimum non-zero eigenvalue of graph Laplacian and the network density.
	\begin{figure}[htbp]
		\centering
		\subfloat[$P_4$ with $\lambda_{2}=0.5858$]{\label{fig4(a)}%为子图加交叉引用	
			\begin{minipage}[b]{0.3\linewidth}
				\centering
				\includegraphics[]{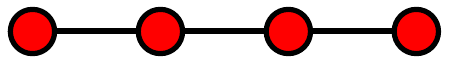}
			\end{minipage}
		}
		\quad
		\subfloat[$S_4$ with $\lambda_{2}=1$]{\label{fig4(b)}%
			\begin{minipage}[b]{0.3\linewidth}
				\centering
				\includegraphics[]{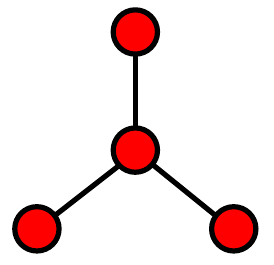}
			\end{minipage}
		}
		\quad
		\subfloat[$C_{1,4}$ with $\lambda_{2}=2$]{\label{fig4(c)}%
			\begin{minipage}[b]{0.3\linewidth}
				\centering
				\includegraphics[]{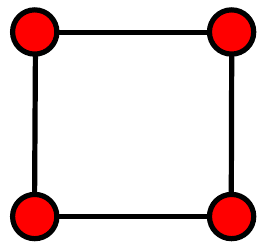}
			\end{minipage}
		}
		\caption{Diagrams of $P_4$, $S_4$, and $C_{1,4}$.}
		\label{fig4}
	\end{figure}
	
	\begin{table}[htbp]
		\centering
		\caption{The analytic expressions of the minimum non-zero eigenvalue and the network density.
			\label{tab-1}}
		\resizebox{0.6\textwidth}{!}{
			\begin{tabular}{cccccc}
				\toprule\addlinespace
				Graph & Value of $\lambda_{2}$ & Network density $d$ & $\lambda_{2}<1$ & $\lambda_{2}=1$& $\lambda_{2}>1$\\
				\addlinespace
				\hline \addlinespace
				$K_n$ & $\lambda_{2}=n$ & $d=1$ &
				\XSolidBrush & \XSolidBrush & $n\geq2$ \\
				\addlinespace
				\multirow{2}{*}{$S_n$} & $\lambda_{2}=1~(n\geq 3)$ & \multirow{2}{*}{$d=\frac{2}{n}$} & \multirow{2}{*}{\XSolidBrush} &
				\multirow{2}{*}{$n\geq3$} & \multirow{2}{*}{$n=2$} \\ 
				&$\lambda_{2}=2~(n=2)$& & & & \\
				\addlinespace
				$P_n$ & $\lambda_{2}=4\sin^2(\dfrac{\pi}{2n})$ & $d=\frac{2}{n}$ & $n>3$ & $n=3$ & $n=2$ \\
				\addlinespace
				$C_{1,n}$  & $\lambda_{2}=3-\dfrac{\sin\frac{3\pi}{n}}{\sin\frac{\pi}{n}}$ & $d=\frac{2}{n-1}$ &$n\geq7$ & $n=6$ & $n\leq5$ \\
				\addlinespace
				$C_{2,n}$  & $\lambda_{2}=5-\dfrac{\sin\frac{5\pi}{n}}{\sin\frac{\pi}{n}}$ & $d=\frac{4}{n-1}$ & $n\geq14$ & \XSolidBrush & $n\leq13$ \\
				\addlinespace
				$C_{3,n}$  & $\lambda_{2}=7-\dfrac{\sin\frac{7\pi}{n}}{\sin\frac{\pi}{n}}$ & $d=\frac{6}{n-1}$ & $n\geq24$ & \XSolidBrush & $n\leq23$ \\
				\bottomrule
		\end{tabular}}
	\end{table}	
	
\begin{figure}[htbp]
		\centering
		\subfloat[The trends of $\|T_1(s)\|_\infty-\|T_2(s)\|_\infty$ for graph $P_4$]{\label{fig5(a)}%为子图加交叉引用	
			\includegraphics[width=0.48\textwidth,height=0.4\textwidth]{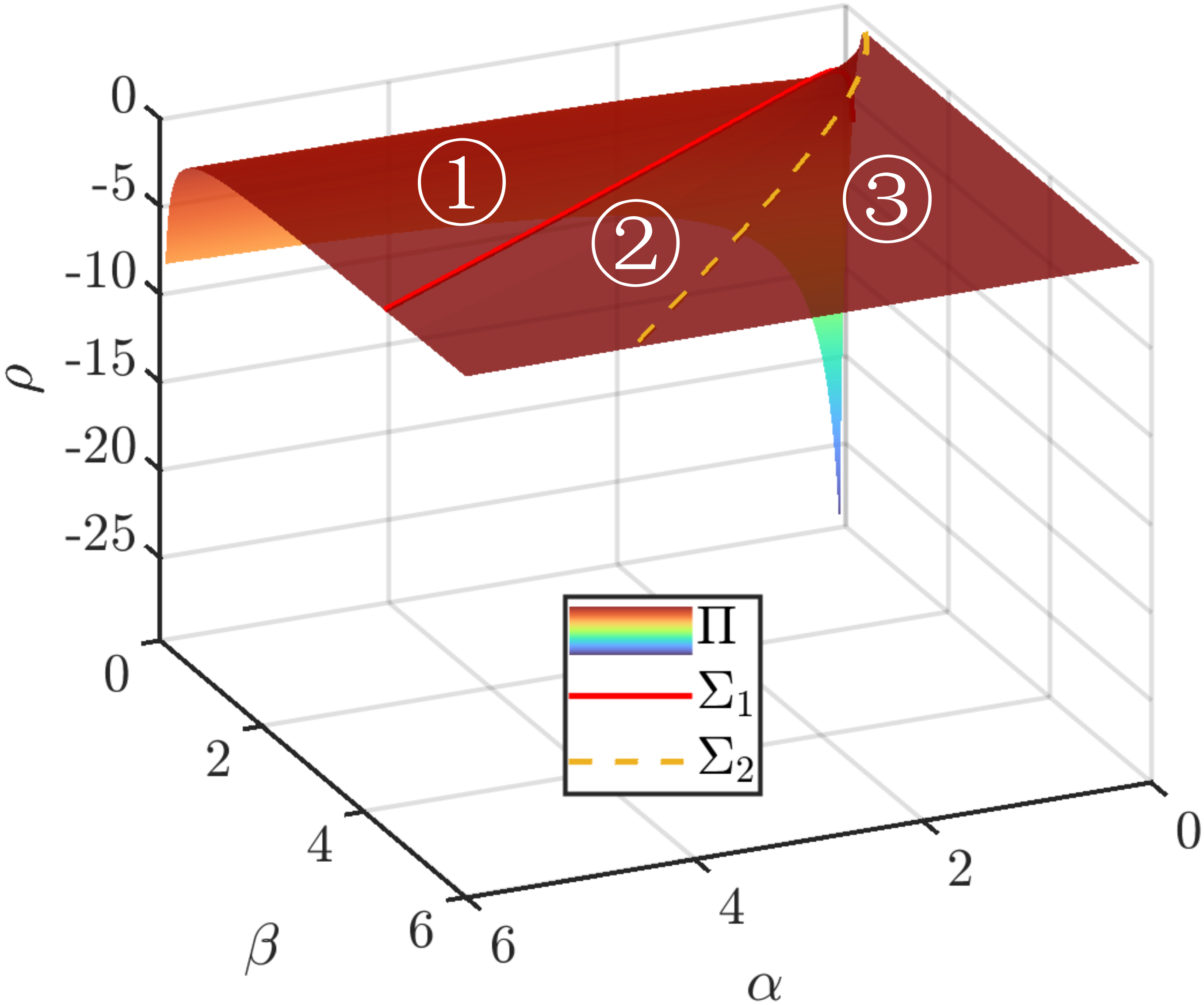}
		}
		\\
		\subfloat[The trends of $\|T_1(s)\|_\infty-\|T_2(s)\|_\infty$ for graph $S_4$]{\label{fig5(b)}%
			\includegraphics[width=0.48\textwidth,height=0.4\textwidth]{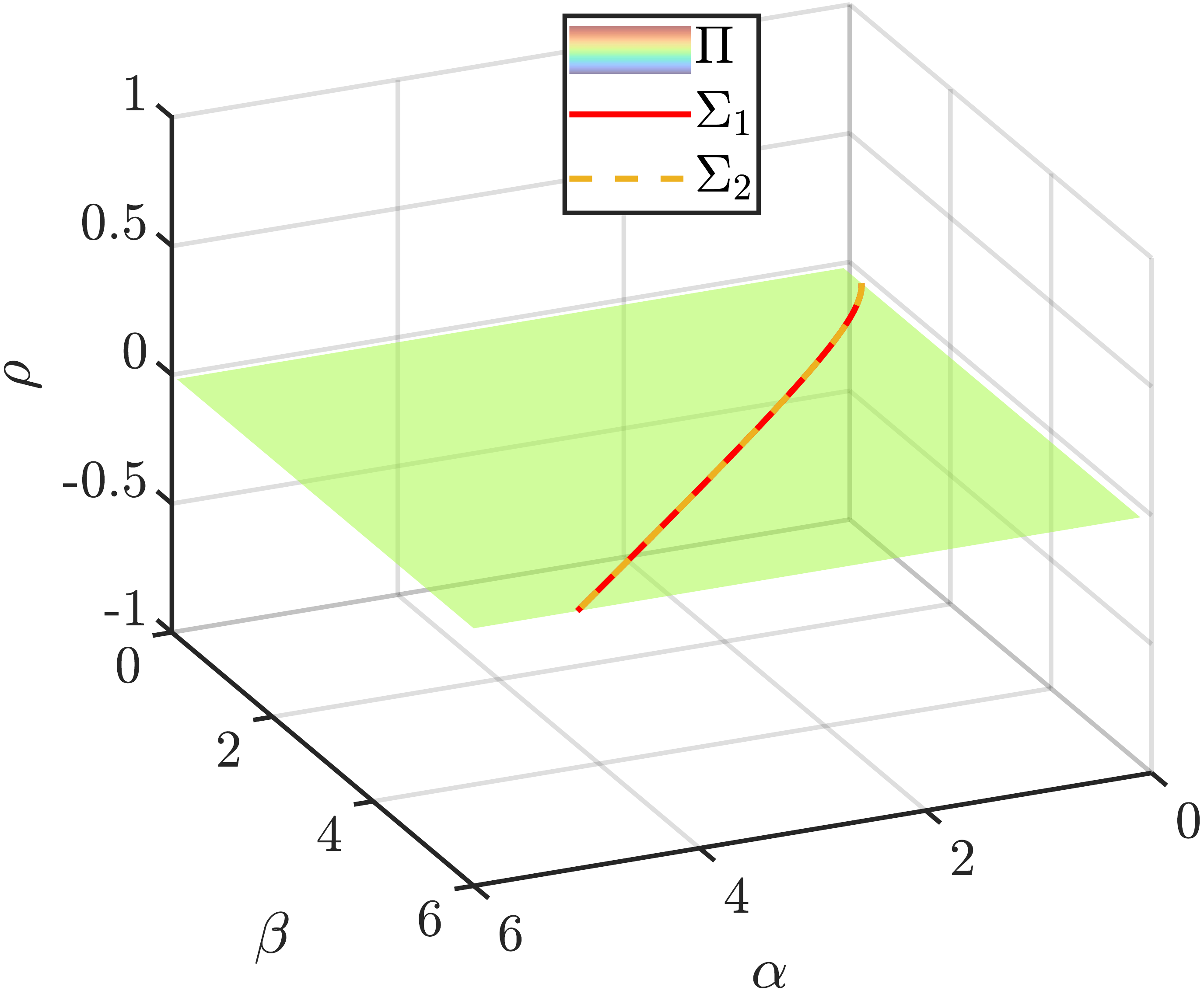}
		}
		\\
		\subfloat[The trends of $\|T_1(s)\|_\infty-\|T_2(s)\|_\infty$ for graph $C_{1,4}$]{\label{fig5(c)}%
			\includegraphics[width=0.48\textwidth,height=0.4\textwidth]{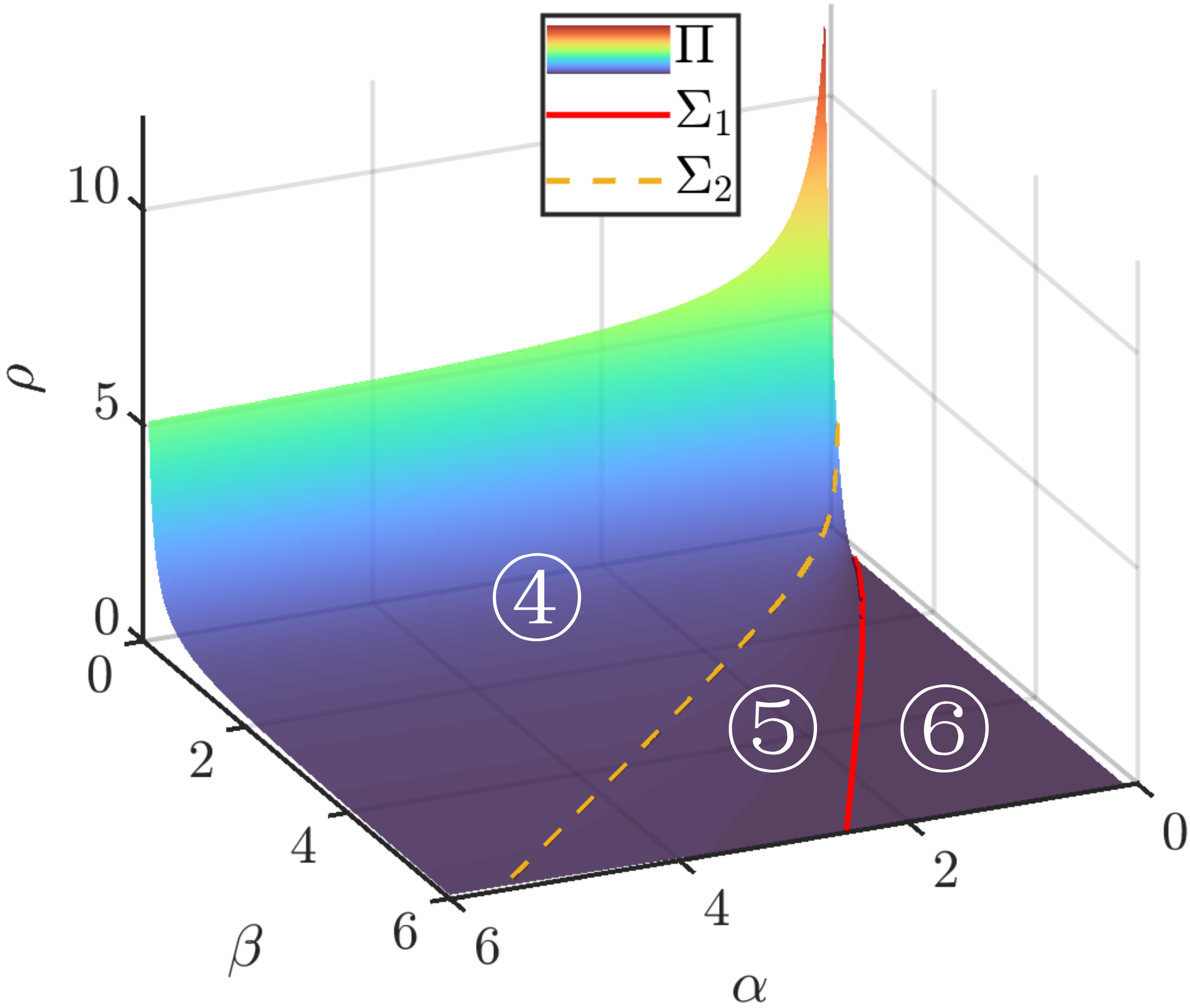}
		}
		\caption{The trends of $\|T_1(s)\|_\infty-\|T_2(s)\|_\infty$ with respect to tunable gains $\alpha$ and $\beta$ for the graphs $P_4$, $S_4$ and $C_{1,4}$.}
		\label{fig5}
	\end{figure}

	%    Then three graphs $P_4$, $S_4$ and $C_{1,4}$ with different $\lambda_{2}$ depicted in Fig. \ref{fig4} are picked to illustrate our graph condition.
	In simulations, we will use three graphs $P_4$, $S_4$ and $C_{1,4}$, whose diagrams are depicted in Fig. \ref{fig4}, to illustrate our findings.
	For each network in Fig. \ref{fig4}, the two-dimensional surface $\Pi:=\Big\{(\alpha,\beta,\rho)\in\mathbb{R}^3~\big|~\alpha\in\mathbb{R}^+,\beta\in\mathbb{R}^+,\rho=\|T_1(s)\|_\infty-\|T_2(s)\|_\infty\Big\}$
	is displayed in Fig. \ref{fig5}, which denotes the trends of difference between $\|T_1(s)\|_\infty$ and $\|T_2(s)\|_\infty$ with respect to tunable gains $\alpha$ and $\beta$. 
	We respectively depict the red solid curve
	$$\Sigma_1:=\Big\{(\alpha,\beta,\rho)\in\Pi~\big|~\beta=\sqrt{(\alpha\lambda_{2})^2+2\alpha\lambda_{2}}\Big\}$$
	and the yellow dash curve
	$$\Sigma_2:=\Bigg\{(\alpha,\beta,\rho)\in\Pi~\bigg|~\beta=\sqrt{\alpha^2+\dfrac{2\alpha}{\lambda_2}}\Bigg\}.$$
	on $\Pi$ to emphasize the critical conditions in \eqref{eq:Th1} and \eqref{eq:Th2}.
	
	As one can see from Fig. \ref{fig5(a)}, the surface $\Pi$ is divided into three regions \ding{172}, \ding{173} and \ding{174} by curves $\Sigma_1$ and $\Sigma_2$, which correspond to the three cases (\uppercase\expandafter{\romannumeral1}), (\uppercase\expandafter{\romannumeral2}) and (\uppercase\expandafter{\romannumeral3}) in the proof of Theorem \ref{theorem:3}, respectively. Obviously, the two regions \ding{172} and \ding{173} are always below the plane $\rho=0$ which means that $\|T_1(s)\|_\infty<\|T_2(s)\|_\infty$.
	While the region \ding{174} completely overlaps the plane $\rho=0$ which implies that $\|T_1(s)\|_\infty=\|T_2(s)\|_\infty$. 
	%    These observations are in accordance with Theorem \ref{theorem:3} which means that the absolute velocity protocol \eqref{eq:1.2} is superior to the relative velocity protocol \eqref{eq:1.3} if $\lambda_{2}<1$.
	According to Definition \ref{Def:3}, these observations tells that the absolute velocity protocol \eqref{eq:1.2} is superior to the relative velocity protocol \eqref{eq:1.3} when $\lambda_2<1$, which are in accordance with the conclusion 1) in Theorem \ref{theorem:3}.
	Likewise, as shown in Fig. \ref{fig5(c)}, the surface $\Pi$ is separated into three regions \ding{175}, \ding{176} and \ding{177} by curves $\Sigma_1$ and $\Sigma_2$. Obviously, bounded by the red solid line, the two regions \ding{175} and \ding{176} are always above the plane $\rho=0$ which means that $\|T_1(s)\|_\infty>\|T_2(s)\|_\infty$.
	While the region \ding{177} completely overlaps the plane $\rho=0$ which implies that $\|T_1(s)\|_\infty=\|T_2(s)\|_\infty$. 
	Therefore, it follows from Definition \ref{Def:3} that the relative velocity protocol \eqref{eq:1.3} surpasses the absolute velocity protocol \eqref{eq:1.2} when $\lambda_{2}>1$, which is consistent with the conclusion 2) in Theorem \ref{theorem:3}. It is observed from Fig. \ref{fig5(b)} that when $\lambda_{2}=1$, then $\|T_1(s)\|_\infty\equiv\|T_2(s)\|_\infty$ because the surface $\Pi$ completely overlaps the plane $\rho=0$. In other words, the absolute velocity protocol \eqref{eq:1.2} performs as well as the relative velocity protocol \eqref{eq:1.3} when $\lambda_{2}=1$, which is consistent with the finding presented in Theorem \ref{theorem:3}.
	
\begin{figure}[!ht]\centering
		\includegraphics[width=0.48\textwidth,height=0.36\textwidth]{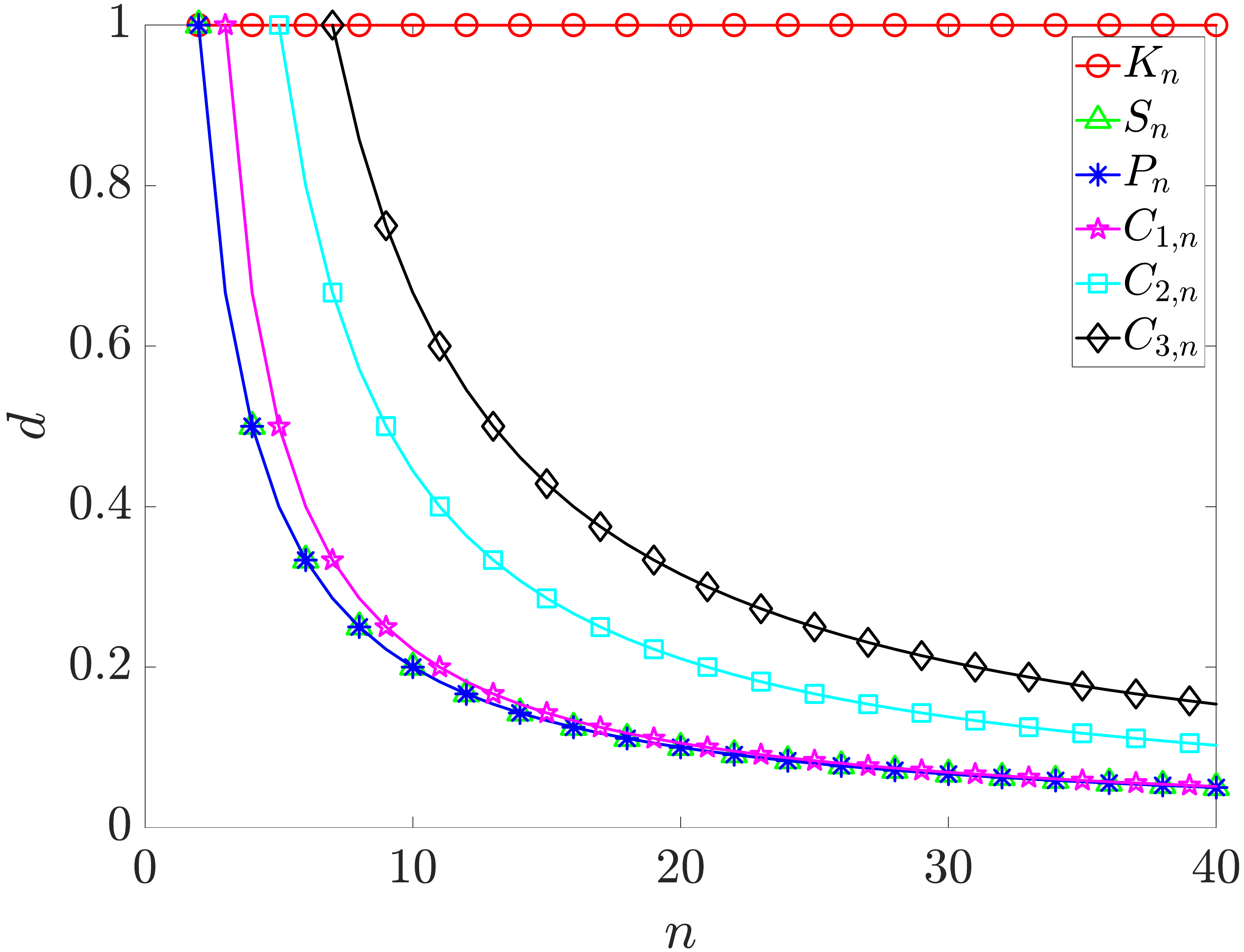}
		\caption{Trends of network density $d$ on various graphs.\label{fig6}}
	\end{figure}

	Furthermore, according to Theorem \ref{theorem:3} and Table \ref{tab-1}, for the undirected complete graph with any number of nodes, the relative velocity protocol is preferable to the absolute velocity protocol. 
	For the path graph with the number of nodes greater than 3, the absolute velocity protocol is superior to the relative one.
	When the star graph whose number of nodes is not less than 3 is considered, both the two protocols have the same anti-disturbance capability.
	For the $2$-regular ring lattice, $4$-regular ring lattice and $6$-regular ring lattice, the absolute velocity protocol outperforms the relative velocity protocol when the number of nodes is not less than 7, 14 and 24, respectively, and the relative velocity protocol is superior to the absolute one when the number of nodes is note greater than 5, 13 and 23, respectively.
	Particularly, for the $2$-regular ring lattice with 6 nodes, both the two protocols own the same anti-disturbance capability.
	Based on these observations, our graph conditions seem to be consistent with the variation of network density $d$ whose trends as increasing number of nodes $n$ is shown in Fig. \ref{fig6}.
	%    Furthermore, in order to obtain better anti-disturbance capability, combining Theorem \ref{theorem:3} with Table \ref{tab-1} yields that: 1) we prefer to select the relative velocity protocol \eqref{eq:1.3} for the undirected complete graph; 2) the absolute velocity protocol \eqref{eq:1.2} is better for the path graph if the size of network are large. Note that both protocols \eqref{eq:1.2} and \eqref{eq:1.3} are equivalent for the star graph when $n\geq 3$. Furthermore, the protocol selection for the regular ring lattice is closely related to the number of neighbors $2k$ and the size of group $n$. More specifically, we are more likely to pick the absolute velocity protocol \eqref{eq:1.2} for a large-scale network. For a certain scale of group, a large $k$ corresponds to select the relative velocity protocol \eqref{eq:1.3} while a small $k$ implies that the absolute velocity protocol \eqref{eq:1.2} is superior.
	
	In Fig. \ref{fig6}, the network density $d$ is diminishing with the growth in number of nodes except the complete graph which retains the highest  density all through. Combining Fig. \ref{fig6} with Theorem \ref{theorem:3} and Table \ref{tab-1} obtains that the absolute velocity protocol \eqref{eq:1.2} can be viewed as the optimal selection when the network density $d$ is not higher than a certain threshold, otherwise the relative velocity protocol \eqref{eq:1.3} is always the best. And the threshold varies from different families of graphs.

	%%%%%%%%%%%%%%%%%%%%%%%%%%%%%%%%%%%%%%%%%%%%%%%%%%%%%%%%%
	\section{Conclusion}\label{Sec:5}
	%%%%%%%%%%%%%%%%%%%%%%%%%%%%%%%%%%%%%%%%%%%%%%%%%%%%%%%%%
	
	%%%%%%%%%%%%%%%%%%%%%%%%%%%%%%%%%%%%%%%%%%%%%%%%%%%%%%%%%
	
	In this paper, we investigated the anti-disturbance capability for the second-order MASs with absolute and relative velocity protocols, respectively, and gave the graph conditions to show which protocol owns better anti-disturbance capability. 
	The anti-disturbance capability was characterized by the \Ltwo gain from the disturbance to the consensus error.
	Firstly, we built the analytic expression of the \Ltwo gain of the MAS with absolute velocity protocol. Then the analytic expression of the \Ltwo gain of the MAS with relative velocity protocol was also established.
	It was shown that both the \Ltwo gains for absolute and relative velocity protocols only depend on the minimum non-zero eigenvalue $\lambda_{2}$ of Laplacian matrix $L$ and the tunable gains $\alpha$ and $\beta$.
	Secondly, based on the analytic expressions of the \Ltwo gain, we put forward the graph conditions for protocol selection for better anti-disturbance capability. It was proved that the absolute velocity protocol is superior to the relative velocity protocol if $\lambda_{2}<1$, and the relative velocity protocol outperforms the absolute one if $\lambda_{2}>1$. Especially, both the two protocols have the same anti-disturbance capability if $\lambda_{2}=1$.
	Moreover, we presented a two-step method for improving anti-disturbance capability. Finally, simulations illustrated the effectiveness of our findings. Although it might be intuitively true that the network density is associated with protocol selection, this fact deserves to be further verified in future works.

	%%%%%%%%%%%%%%%%%%%%%%%%%%%%%%%%%%%%%%%%%%%%%%%%%%%%%%%%%%%%%%%%%%%%%%

	%% The Appendices part is started with the command \appendix;
	%% appendix sections are then done as normal sections
	%% \appendix
	
	%% \section{}
	%% \label{}
	
	%% If you have bibdatabase file and want bibtex to generate the
	%% bibitems, please use
	%%
	%%  \bibliographystyle{elsarticle-num} 
	%%  \bibliography{<your bibdatabase>}
	
	%% else use the following coding to input the bibitems directly in the
	%% TeX file.
	
	%\begin{thebibliography}{00}
	%
	%%% \bibitem{label}
	%%% Text of bibliographic item
	%
	%\bibitem{}
	%
	%\end{thebibliography}
	
	\bibliography{references}
\end{document}